# A Proactive Malicious Software Identification Approach for Digital Forensic Examiners

**Muhammad Ali[1] , Stavros Shiaeles [1], Nathan Clarke[1] and Dimitrios Kontogeorgis[2]**

[1] Centre for Security, Communications, and Networks Research (CSCAN),
School of Computing and Mathematics,
Plymouth University, UK

[2]Open University of Cyprus,
School of Applied Science,
Latsia, Nicosia, Cyprus

## Abstract

Digital investigators often get involved with cases which seemingly point the responsibility to the person to which the computer belongs, but after a thorough examination malware is proven to be the cause, causing loss of precious time. Whilst Anti-Virus (AV) software can assist the investigator in identifying the presence of malware, with the increase in zero-day attacks and errors that exist in AV tools, this is something that cannot be relied upon. The aim of this paper is to investigate the behavior of malware upon various Windows operating system versions in order to determine and correlate the relationship between malicious software and OS artifacts. This will enable an investigator to be more efficient in identifying the presence of new malware and provide a starting point for further investigation.

The study analyzed several versions of the Windows operating systems (Windows 7, 8.1 and 10) and monitored the interaction of 90 samples of malware across three categories of the most prevalent (Trojan, Worm, and Bot) and 90 benign samples through the Windows Registry. Analysis of the interactions has provided a rich source of knowledge about how various forms of malware interact with key areas of the Registry. Using this knowledge, the study sought to develop an approach to predict the presence and type of malware present through an analysis of the Registry. To this end, different classifiers such as Neural Network, Random forest, Decision tree, Boosted tree and Logistic Regression were tested. It was observed that Boosted tree was resulting in a correct classification of over 72% – providing the investigator with a simple approach to determining which type of malware might be present independent and faster than an Antivirus. The modeling of these findings and their integration in an application or forensic analysis within an existing tool would be useful for digital forensic investigators.

## 1. Introduction

As malware evolves and becomes more complex, malicious attackers are able to adapt their behavior depending on the system they wish to infect. Malicious software can only be revealed after the recognition of specific factors of the system and the combination of many parameters and conditions. For example, a particular malware might reveal it’s behavior when installing on a Windows 7 platform or when specific software is installed on the victim's computer (for example a PDF Reader) and to remain totally inactive in any other situation. Similarly, it can reveal a part of his behavior, while parts of the functionality remain hidden until certain conditions that will cause additional activity. Attempts have been made to uncover malware activated behavior (Moser et al., 2007; Brumley et al., 2008) but it has also shown for it to be possible to trick such analyzers (Sharif et al., 2008). During the examination of a case, the possibility always exists that digital evidence or criminal activity is the result of malware activity. It could be that the owner of a computer system is unjustly suspected due to the presence of malicious software. Therefore, in each case prior to the recording of evidence, a thorough investigation for the presence of malware should be undertaken. Traditionally, this is achieved using one or more Anti-Virus (AV) systems. However, weaknesses in AV technology and the increasing presence of zero-day vulnerabilities make them less than full-proof.

Many researchers have conducted studies to find digital artifacts on the Windows operating system, including earlier versions of Windows, were analyzed, such as Vista (Purcell and Lang, 2008), 7 (Thomas et al., 2013) and 8 (Stormo, 2013). Further research has also been undertaken on the study of specific operating regions such as the Registry (Mee et al., 2006), volatile memory (Schuster, 2006; Dolan-Gavitt, 2008; Thomas et al., 2013; Shanks, 2014), USB devices (Carvey and Altheide, 2005; Collie 2013) and the file system (Carrier, 2005; Carvey, 2009; Malicious-streams, 2014). In this research work a comparative study of three core Windows OSs (Windows 7, 8.1 and 10) is undertaken in order to study whether the version of OS has an impact over the behavior and performance of malicious software. This will provide digital forensics analysts with invaluable help, as they will have a guide for the locations to which are expected to have digital evidence. With a targeted investigation at specific locations, it is possible to identify whether a system is infected with malware or not. The paper also develops and evaluates an approach to automatically predict which type of malware is present. This allows forensic examiners to more quickly and reliably identify the presence and type of malware.

## 2. Background and Related Work

The detection of malware through an analysis of unknown executables is not a new problem. Consequently, many solutions already exist. These solutions can be divided into two categories: static and dynamic analysis.

### 2.1. Static Analysis

In static analysis, an incident response team analyzes the code or the structure of a program to determine the functionality without running the program (Sikorski and Honig, 2012). First steps include the use of all available anti-virus programs. This could give information to a known malware for which signature is available, saving valuable time in the process. A major disadvantage of this technique is the dependence upon the detection of the virus based largely on file signatures. Malicious code developers can easily change the code in order to avoid detection (Dalziel, 2014). Another technique used in the static analysis is binary code disassembling, which converts the binary code into an assembly and then analytical techniques control data flow resulting in a report of the running program. A series of binary code analysis techniques (Christodorescu and Jha, 2003; Kruegel, et al. 2004; Christodorescu et al., 2005) have been presented for the detection of different types of malware. The advantage of static analysis is that it’s carried out quickly and that can cover the entire application code. Whilst, there is rich literature on static analysis techniques, which indicates that many problems can be tackled well in practice due to predictability, often this is because it is being applied to real applications rather than malware. Unfortunately, since malware is directly created by cyber criminals it can be deliberately crafted so that it is difficult to analyze. Specifically, the attacker can make use of technical binary obfuscation to prevent both the disassembly of the code and analysis, methods which are used by static analysis techniques.

## 2.2. Dynamic Analysis

The dynamic analysis techniques of malware behavior characterized by the analysis of the actual instructions of a program or the results it brings the program to the operating system. Compared with the static approach, dynamic analysis is less susceptible to various code obfuscation techniques (Moser et al., 2007). Christodorescu et al., (2007) introduce the specifications of malware using data flows between the system calls. They found the actual relationships between system calls are difficult to overcome with random system calls. Since then, this knowledge of malicious software has been widely used in malware analysis tasks such as extraction of distinct malware functions, mining the difference between malware behaviour and benign behaviour of the program (Fredrikson et al., 2010), determining malware families in which samples are sharing common functions (Bayer et al., 2009; Babić et al., 2011; Park et al., 2013) and to detect malicious behaviour ( Lippmann and Clark., 2008; Kolbitsch et al., 2009; Bayer et al., 2010). Another method uses a representative audience behavior chart for all samples of malware in a family, instead of a behavior chart per case. The proposed approach is valid and effective since most new malware variants are from known families (Gordon, 1997; Park et al., 2010; Vlachos et al., 2012). Despite various metamorphic and polymorphic blackouts, samples of malicious software within the same family tend to reveal similar malicious behavior (Lindorfer et al., 2012).

The most popular method of analyzing the malware operation in a safe way is to use sandbox technology. The sandbox is running as a separate system, contains the untrusted program and prevents any action from accessing the real network and often provides network services for malware in a form of "black hole." If the untrusted program makes a DNS request, for example, the sandbox will answer the question, usually with 127.0.0.1 (loopback).

Since the spread of metamorphic and polymorphic viruses, dynamic analysis of malware has been established as an effective approach to understanding and classifying malware by observing the execution of malware samples in quarantine environment (Willems et al., 2007; Egele et al., 2012). The interaction between the execution of the malicious sample and operating system allows dynamic malware analysis systems to collect those behavioral characteristics that help shape technical defense.

A problem that was found in modern viruses is that the malicious code is often equipped with detection routines that check for the presence of a virtual machine or a simulated operating system environment. When such an environment is detected, the malware modifies its behavior and the analysis yields incorrect results or even worst, the malware stops to function making analysis impossible. Moreover, some malware also checks for software (even material) having breakpoints to detect whether the program is running in a debug program. In order to bypass the aforementioned problems, the analysis environment should be invisible to malicious code, comprehensive and cover all aspects of the interaction of an environmental program.

Efforts to investigate the possible prevention of malware incidents have prompted earlier studies on malicious codes (Balthrop et al., 2004; Costa et al., 2005; Zou et al., 2007). Current research in cybersecurity focuses on the characterization and modeling of specific attacks, with the aim of understanding the mechanisms of penetration, detection, and response. As cyber threats are increased both in number and in complexity, it has increased the interest in infectious malware (Liu et al., 2016). In theory, one of the interesting issues is the creation of reliable mathematical models that can be applied to effectively describe and forecast the evolution of malicious computer software. Since the spread of malicious code is similar to the biological epidemics (Vespignani, 2005), some epidemiological models have been employed to study the behavior of malicious software (Cheng et al., 2011; Li et al., 2014; Mishra and Pandey, 2014; Misra et al., 2014; Shukla et al., 2014). In addition, new strategies and methodologies necessary to prevent invasions and addressing their effects (Gil et al., 2014).

## 3. Experimental Methodology

The purpose of this research was to examine in a dead-box mode the impact that malware has on the Windows Registry. Furthermore, the research sought to understand what differences exist in differing versions of the OS. To this end, 90 samples of both malware (split between Trojans, worms, and bot) and cleanware were selected to provide a robust and comprehensive analysis. The Microsoft's Windows operating systems was focussed upon as, it is still the prominent OS in use today (NetMarketShare, 2016).

### 3.1 Virtual lab

The analysis laboratory consists of two testbeds. The first one is running locally on a host machine with a CPU Intel Core i7-4790, 16 GB RAM and Windows 10 Pro is the bare metal configuration. In this machine we installed VMware Workstation Pro 12 (VMware, 2016) and virtual Ubuntu 16.04 operating system which hosted the Cuckoo Sandbox. Ubuntu compatibility of the Cuckoo Sandbox (Oktavianto and Muhardianto, 2013) is excellent and has been used by other investigators (Shanks, 2014) for the same purpose.

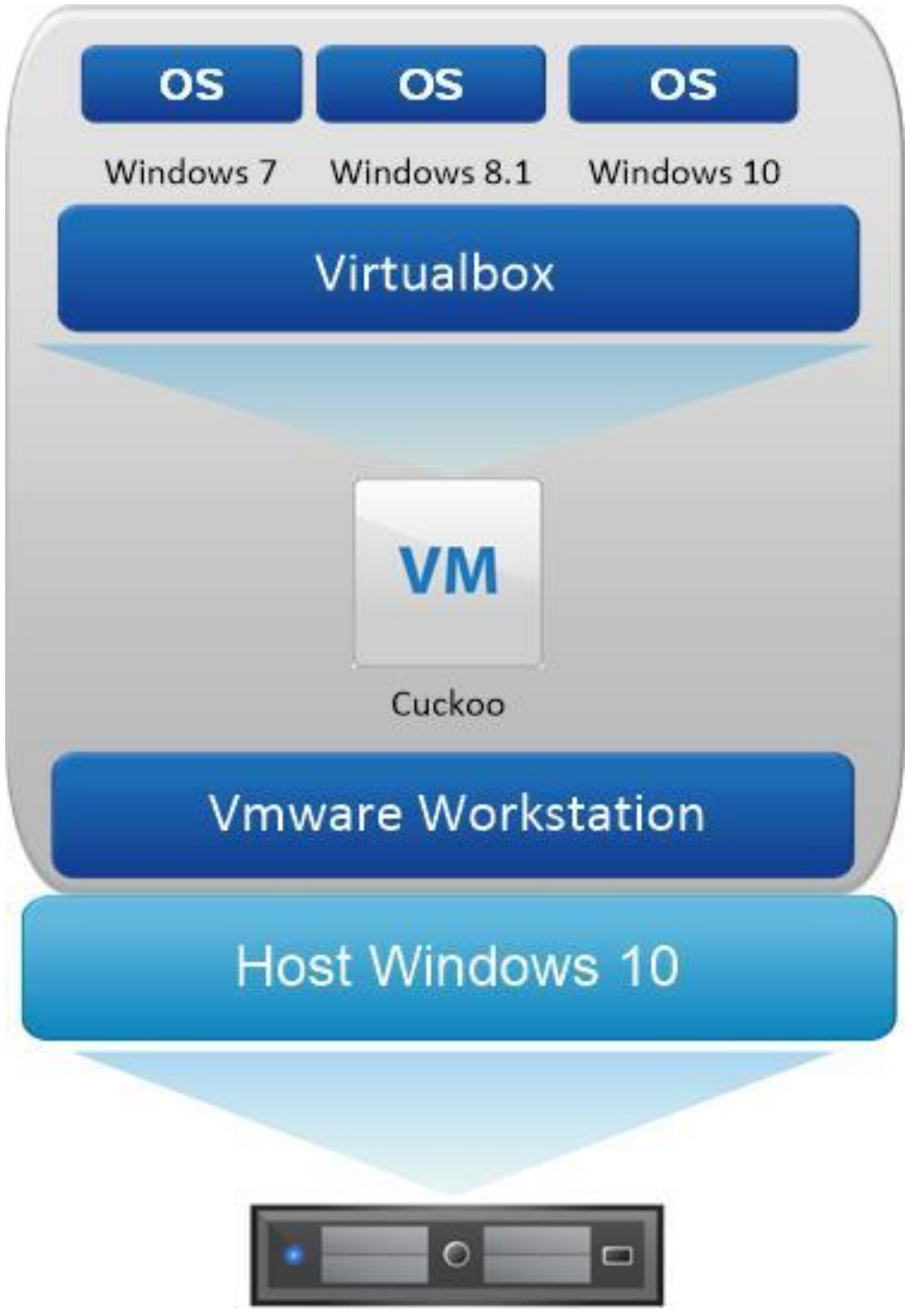

**Figure 1: Virtual lab architect**

Furthermore, three virtual machines were created with the following characteristics: CPU a core of the Intel Core i7-4790, 2 GB RAM and Operating System Windows. In each of these three virtual machines installed a different version of Windows, specifically 7, 8.1 and 10. In order to communicate the Cuckoo with each operating to be tested, one of Cuckoo network card and unique virtual

Windows machine card, connected to a virtual isolated network (192.168.56.0/24) as shown in Figure 2.

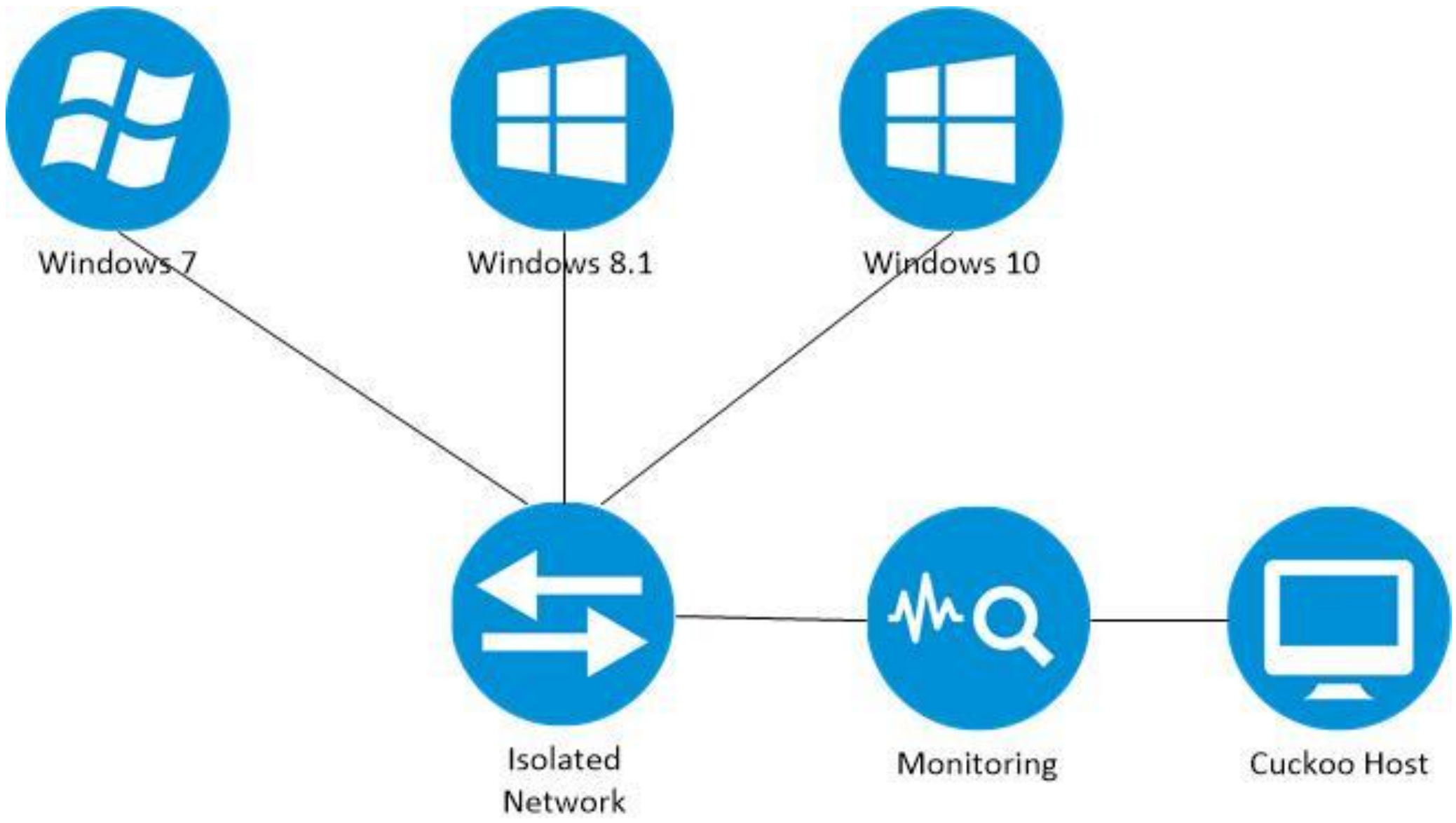

**Figure 2: Virtual network configuration**

The script (in Python) and the libraries are Cuckoo's important components (Hosmer, 2014). The system consists of a server where the Cuckoo software installed, a virtualization program such as Oracle VirtualBox (Dash, 2013) and the Cuckoo agent that runs in a virtual machine. A guest system is a virtual machine that is running one of the Windows versions. This virtual machine has turned off the User Account Control (UAC), automatic updates and firewall.

A methodology for the understanding malicious code is based upon sandboxing (Greamo and Ghosh, 2011). In simple terms, the process involves the execution of arbitrary code in a controlled manner that allows direct observation of results. By documenting evidence, such as open ports, registry keys, IP addresses, file incorporated and domain names, a team can gain information on the regular opponent.

The isolated environment allows the sample to run without adversely affecting the system host computer or the quest. Once the desired state of the system has accomplished, a system snapshot is taken. This snapshot can be used to restore the system to a known clean state after the sample is analyzed. A command line interface used to all commands executed within Cuckoo Sandbox.

The second testbed was hosted on cloud and we used two cloud sandboxes to withdraw as much information as possible in order to find more unique registry hives from the malware and cleanware. For this experiment we utilised an Agentless (VMRay Analyzer) and AI-based (SNDBOX) sandbox. VMRay Analyser as aforementioned is an agentless sandbox cloud solution and the reason choosing this platform is that some sophisticate malware usually monitor the running environment and to prevent their discovery they usually stop their execution which provides insignificant features to the analysis (Ali et al., 2018). SNDBOX applies an invisible kernel mode agent and AI to offer the next generation Sandbox, extending the individual capabilities and expertise of security and research teams through AI, dynamic analysis and network mapping. It is Located between the User mode and Kernel mode, SNDBOX's invisible agent deceives malware into executing its full range of intended functionality, revealing its true malicious nature, intent, and capabilities ("SNDBOX," 2019).

### 3.2 Standardized naming scheme for malware

Security analysts and researchers from different AV companies in 1991 developed a standardized naming scheme for malware known as Computer Antivirus Research Organization (CARO) (“A New Virus Naming Convention (1991) - CARO - Computer Antivirus Research Organization,” n.d.). The philosophy behind the development of this standard is to remove the confusion among the users and AV-Vendors by having a common standard or syntax for naming malware. The generic form of this format is mentioned below

<malware type>://<platform>/<family name>.<group name>.<infective length>.<sub-variant><devolution><modifiers>

In the string above only family name is compulsory and the rest of the fields are optional. Although most companies claim that they follow the CARO scheme, but in practice only Microsoft is using this convention in their AV software for MSE or the Win8 version and windows defender etc.

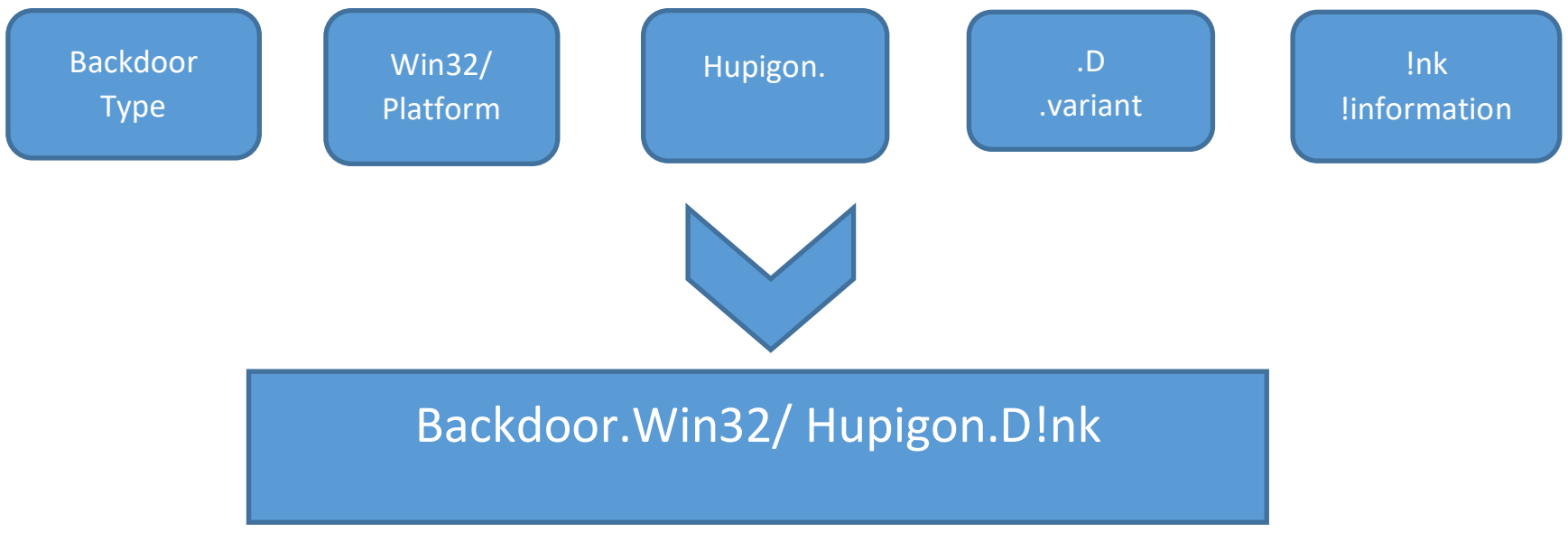

Figure 3: CARO malware naming scheme (“Malware names | Microsoft Docs,” n.d.)

In this research work we used the CARO naming scheme to name malware as shown in section 3.3

### 3.3 Malware samples used

Often, malware investigators have to deal new threats and unknown executable. In some cases, there are scenarios where you can handle malware that already knows its name and is classified, e.g. for research purposes as in this work. To analyze such malicious software, there are many places where one researcher can collect known samples. The Lenny Zeltser, who is the head of the private SANS Institute (SANS, 2016), recommends several free resources on his website (Zeltser, 2016). The samples of malicious code used in this research were taken from Malware.lu (Malware.lu, 2016), Virussign (VirusSign, 2016), Vx Heaven (VxHeaven, 2016), Malekal (Malekal, 2016) and MalwareTips (MalwareTips, 2016). Ninety samples of malicious code were selected, 20 Trojan, 20 Worms, and 20 Bot. These three malware families were selected (table 1) as they are the main categories that are detected more often (Veracode, 2012; AlienVault, 2013; Malwaretruth, 2016).

**Table 1: The malware samples that were used**

| | Category | Virus name | SPY-STEAL DATA | C&C | BACKDOOR | STEALTH |
|---|---|---|---|---|---|---|
| **1** | Trojan | Trojan-Spy.Win32.Zbot.wijf | X | X | | |
| **2** | Trojan | Trojan.GenericKD.3015891 | | X | | |
| **3** | Trojan | Trojan.GenericKD.3015909 | | | X | |
| **4** | Trojan | Trojan/Win32.Yakes | | | X | |
| **5** | Trojan | Trojan.GenericKD.3016131 | | | X | |
| **6** | Trojan | Trojan/W32.KRBanker | | X | | |
| **7** | Trojan | Trojan-Spy.Win32.FlyStudio.ij | | | X | X |
| **8** | Trojan | Trojan-Dropper.Win32.Injector.nyds | | | X | |
| **9** | Trojan | Trojan.Zboter | X | X | | |
| **10** | Trojan | Trojan-Spy.Win32.Recam.yue | X | X | | |
| **11** | Trojan | Trojan. Tesla!1.A322 | | X | | X |
| **12** | Trojan | Trojan.Win32.Waldek.cbp | | | X | |
| **13** | Trojan | Trojan.Win32.Waldek.cbm | | | X | |
| **14** | Trojan | Trojan.Win32.Dridex.v | X | X | | X |
| **15** | Trojan | Trojan.Win32.Tepfer.psxezj | X | X | | |
| **16** | Trojan | Trojan.Win32.Yakes.owmp | | | X | |
| **17** | Trojan | Trojan.Win32.KeyLogger.auqd | X | | | |
| **18** | Trojan | Trojan.GenericKD.3023498 | | X | | |
| **19** | Trojan | Trojan.Generic.8742442 | X | X | | X |
| **20** | Trojan | Trojan.Generic.7738292 | X | | | |
| **21** | Trojan | Trojan.Generic.<br>AAA._xeDropperSpywareTrojan | X | X | | |
| **22** | Trojan | Trojan.Generic .Badi | X | X | | X |
| **23** | Trojan | Trojan.Win32.CretClient.exe | | | | |
| **24** | Trojan | Trojan.Generic .InstallBC201401 | | X | | |
| **25** | | Trojan.Generic.pony | X | | | |
| **26** | | Trojan.Generic.Potao_Droppers<br>wdecoy | | | | X |
| **27** | | Trojan.Win32.zeus | X | X | X | X |
| **28** | Trojan | Trojan.Generic.kotbjxfkzeq | X | | | |
| **29** | Trojan | Trojan.Generic.Locky | X | X | | X |
| **30** | Trojan | Trojan. Win32.njRAT.exe | X | X | | X |
| **31** | Trojan | Trojan.Generic.pafish | X | | | |
| **32** | Trojan | Trojan.Win32win32.duqu | | | | |
| **33** | Trojan | Trojan.Generic.Cerber.exe | | | | X |
| **34** | Trojan | Trojan. Win32Mole.exe | | | | X |
| **35** | Trojan | Trojan. Win32.Spora.exe | | | | X |
| **36** | Trojan | Trojan.Win32GrandCrab-01.exe | | | | X |
| **37** | Trojan | Trojan. Win32.Delf.xo | | | X | |
| **38** | Trojan | Trojan. Win32.DarkTequila.exe | | | | |
| **39** | Trojan | Trojan. Win32.psiphon.exe | X | | | |
| **40** | Trojan | Trojan.Generic.yigzwl | | | | X |
| **41** | Trojan | Trojan.Generic.Vcffipzmnipbxzdl | | | | X |
| **42** | Worm | Win32.Gamarue | X | X | | X |
| **43** | Worm | W32.Cridex.A.worm | X | X | | X |
| **44** | Worm | Worm.VBS.Agent | | X | | |
| **45** | Worm | Worm.Win32.3DStars | | | X | X |
| **46** | Worm | Worm.Generic3.PEM | | | X | |
| **47** | Worm | Worm.Win32.Mira.A | X | | | |
| **48** | Worm | Worm.Generic2.CMVO | X | | | |
| **49** | Worm | Worm.Win32.Cake | | | | X |
| **50** | Worm | Worm.Win32.Fever | X | | | X |
| **51** | Worm | Worm.Win32.Monkey.exe | X | | | |
| **52** | Worm | Worm.Win32.Mydoom.a.exe | | | X | X |
| **53** | Worm | Worm.Win32.Pikachu.exe | X | | | |
| **54** | Worm | Worm.Win32.Postman.exe | | | | X |
| **55** | Worm | Worm.Win32.Sharpei.a.exe | | | | X |
| **56** | Worm | Worm.Win32.Silver.exe | | | | X |
| **57** | Worm | Worm.Win32.Sobig.exe | X | X | | |
| **58** | Worm | Worm.KOOBFCE.SMC | X | | X | |
| **59** | Worm | W32/Wabot | X | X | | |
| **60** | Worm | Worm.vid.exe | X | | | |
| **61** | Worm | Email-Worm.Win32.Mydoom.l | | | X | X |
| **62** | Worm | Email-Worm.Win32.Naked | X | | | |
| **63** | Worm | Worm.Christmas-wishes.doc | X | | | |
| **64** | Worm | Win32.WannaCry.EXE | X | | X | X |
| **65** | Worm | Win32.F7F105F9.exe | | | | |
| **66** | Worm | Win32.2tetup.exe | X | | | |
| **67** | Worm | Win32.GrandCrab-01.exe | X | | | |
| **68** | Worm | Win32.GlobeImposter.exe | X | | | |
| **69** | Botnet | Win32.Lolbot.aoi | | | X | |
| **70** | Botnet | WORM/IrcBot.tlq | X | X | X | |
| **71** | Botnet | W32.Jorik_Lolbot.O!tr | X | | X | |
| **72** | Botnet | Win32.SdBot.aamk | X | X | X | |
| **73** | Botnet | W32.ZBot.42352 | X | | X | X |
| **74** | Botnet | Win32.Jorik.SdBot.e | | | X | |
| **75** | Botnet | MSIL.NanoBot.ibh | | | X | |
| **76** | Botnet | Win32.Zbot.vtii | X | | X | X |
| **77** | Botnet | Win32.Ngrbot.anak | | | | X |
| **78** | Botnet | Win32.Alinaos.G | X | X | | |
| **79** | Botnet | GenericKD.2143403 | | | X | |
| **80** | Botnet | Win32/ChkBot.A | | | X | |
| **81** | Botnet | MSIL/Lizarbot.A | X | X | X | |
| **82** | Botnet | Win32.Jorik.Lolbot.f | X | X | X | |
| **83** | Botnet | Win32.Zbot.sbdj | X | | X | X |
| **84** | Botnet | MSIL.NanoBot.bi | X | | X | |
| **85** | Botnet | Win32.Ngrbot.uyk | | | | X |
| **86** | Botnet | Win32.Boht.qo | | X | X | |
| **87** | Botnet | W32/Zbot.AJJU!tr | X | | X | X |
| **88** | Botnet | Win32.VBInject | | | | X |
| **89** | Botnet | Trickbot | | X | | |
| **90** | Botnet | obfuscated.js | | | X | |

## 3.4 Clean samples used

As we wanted to differentiate the clean registry hives from malicious hives cleanware samples such as chrome, teamviewer, skype etc were collected and analysed. During the analysis we also give emphasis to collect system changes along with register hives in order to have more information. The table below shows the samples used as well as the type of each sample.

**Table 2: The cleanware samples that were used**

| | Category | Sample name | Type |
|---|---|---|---|
| **1** | Normal | grammarlyaddinsetup.pe32 | Application software plug |
| **2** | Normal | Poweriso6-x64. | Executable |
| **3** | Normal | Vlc-2-2-1-win32 | Executable |
| **4** | Normal | Wireshark-win64-2.6.5 | Executable |
| **5** | Normal | ProtonVPN.exe | Executable |
| **6** | Normal | Notepad.exe | Executable |
| **7** | Normal | McAfeeWebAdvisor.exe | Executable |
| **8** | Normal | Putty2.exe | Executable |
| **9** | Normal | FTPDesktopClient.exe | Executable |
| **10** | Normal | SQLiteStudio-3.2.1.exe | Executable |
| 11 | Normal | KeePass-2.40-Setup | Executable |
| 12 | Normal | LinuxLiveUSB Creator 2.9.4.exe | Executable |
| 13 | Normal | flashplayer32_install.exe | Executable |
| 14 | Normal | Firefox Setup 14.0.1 | Executable |
| 15 | Normal | 7za.EXE | Executable |
| 16 | Normal | GoogleUpdateSetup.exe | Executable |
| 17 | Normal | Epson512523eu.exe | Executable |
| 18 | Normal | Microsoft-Toolkit.exe | Executable |
| 19 | Normal | Googlewebdesigner_win.exe | Executable |
| 20 | Normal | PDFSAM_Installer.exe | Executable |
| 21 | Normal | FoxitReader_Setup.exe | Executable |
| 22 | Normal | TeamViewer_Setup.exe | Executable |
| 23 | Normal | Internet.Download.Manager.exe | Executable |
| 24 | Normal | TrueCrypt.exe | Executable |
| 25 | Normal | SkypeSetup.exe | Executable |
| 26 | Normal | HottNotes4.1Setup.exe | Executable |
| 27 | Normal | TorchSetup | Executable |
| 28 | Normal | GitHubDesktopSetup | Executable |
| 29 | Normal | Nektar Bolt v1.0 CE.exe | Executable |
| 30 | Normal | ForkInstaller.exe | Executable |
| 31 | Normal | hashcat32.exe | Executable |
| 32 | Normal | AdobePatchInstaller.exe | Executable |
| 33 | Normal | TWUploader.exe | Executable |
| 34 | Normal | vmnat.exe | Executable |
| 35 | Normal | SenseDriver.exe | Executable |
| 36 | Normal | ISSetup.dll | DLL |
| 37 | Normal | SrvCtl.dll | Executable |
| 38 | Normal | panfinder.exe | Executable |
| 39 | Normal | strings.exe | Executable |
| 40 | Normal | procexp.exe | Executable |
| 41 | Normal | cbhqgi.vbs | vbs |
| 42 | Normal | acc.exe | Executable |
| 43 | Normal | KutoolsforExcelSetup.exe | Executable |
| 44 | Normal | DTools.exe | Executable |
| 45 | Normal | winsdk_web.exe | Executable |
| 46 | Normal | ClipboardHistory.exe | Executable |
| 47 | Normal | MEGAsync.exe | Executable |
| 48 | Normal | AnyDesk.exe | Executable |
| 49 | Normal | npp.7.6.Installer.exe | Executable |
| 50 | Normal | CVHP.exe | Executable |
| 51 | Normal | WinSCP-5.13.6-Setup.exe | Executable |
| 52 | Normal | coreftplite64.exe | Executable |
| 53 | Normal | eagleget_setup.exe | Executable |
| 54 | Normal | NetAssemblyInfo.exe | Executable |
| 55 | Normal | Morgan Spencer.htm | htm |
| 56 | Normal | fdminst-lite.exe | Executable |
| 57 | Normal | sigcheck.exe | Executable |
| 58 | Normal | RBInternetEncodings500.dll | DLL |
| 59 | Normal | cryptolibcps-5.0.43.exe | Executable |
| 60 | Normal | Trustlook | PDF |
| 61 | Normal | shell.hta | Executable |
| 62 | Normal | rufus-usb-3-3.exe | Executable |
| 63 | Normal | photosync_setup.exe | Executable |
| 64 | Normal | Home Sweet Home 2 - Kitchens and Baths.exe | Executable |
| 65 | Normal | ThrottleStop.exe | Executable |
| 66 | Normal | Portal.2.incl.upd30-NSIS.exe | Executable |
| 67 | Normal | libeay32.dll | Executable |
| 68 | Normal | PwDump7.exe | Executable |
| 69 | Normal | UaInstall-7.0.6-4.msi | MSI |
| 70 | Normal | TP8-2019.exe | Executable |
| 71 | Normal | tlscntr.exe | Executable |
| 72 | Normal | cccredmgr.exe | Executable |
| 73 | Normal | fdminst-lite.exe | Executable |
| 74 | Normal | HoMM3_HD_Latest.exe | Executable |
| 75 | Normal | ILSpy.exe | Executable |
| 76 | Normal | AnyDesk.exe | Executable |
| 77 | Normal | vs_community__1072350829.1545770560.exe | Executable |
| 78 | Normal | winsdk_web.exe | Executable |
| 79 | Normal | KutoolsforExcelSetup.exe | Executable |
| 80 | Normal | acc.exe | Executable |
| 81 | Normal | cbhqgi.vbs | VBS |
| 82 | Normal | PDFsam_Basic3_3_Installer.exe | Executable |
| 83 | Normal | A_info.pdf | PDF |
| 84 | Normal | Angry Birds.exe | Executable |
| 85 | Normal | aspcmd.msi | MSI |
| 86 | Normal | Research_Paper1.pdf | PDF |
| 87 | Normal | SupportAssistLauncher.exe | Executable |
| 88 | Normal | meda-mp3-joiner-install.exe | Executable |
| 89 | Normal | AutoCopyFiles.exe | Executable |
| 90 | Normal | soffice.exe | Executable |

## 3.5 Sandbox Analysis Procedure

To export information from the samples we performed experiments in three different environments. In the first experiment, Cuckoo was utilised for behavior analysis of malware of files mentioned in section 3.3. and 3.4 above. For each analysis request, a separate subfolder containing all the reports is produced with raw logs, .pcap files, images and any other information obtained during the analysis. Using the Cuckoo as the main malicious software analysis tool, each sample was studied in three different software environments (Windows 7, 8.1 and 10) and the results of the analysis are stored in a suitable form for further study and analysis.

The program sends the sample to the virtual machine that we have selected in the settings file. When injection of the sample into the operating system has completed successfully, Cuckoo monitors all system activity and records it. Once the analysis of the virtual machine is terminated, the .html file with the report of the analysis is created.

The second experiment was completed in two different cloud sandboxes named VMRay analyzer and SNDBOX. In both these sandboxes, benign and malicious samples of section 3.3. and 3.4 respective were executed. For each analysis request, a separate subfolder containing all the reports is produced, the raw logs, .pcap files, images, JSON and any other information obtained during the analysis.

3.5.1 Dataset preparation

Data constitute the input/output variables required to make a prediction. Usually, data comes in two forms either structured or unstructured data. In this research we have taken structure data which implies that data are defined and properly labeled. In order to label data VIT and Virus Total reputation scoring were introduced, to categorize samples as malicious and benign. VirusTotal inspects items with over 70 antivirus scanners database along with URL/domain blacklisting services, in addition to a myriad of tools to extract signals from the studied ("VirusTotal," n.d.). Cuckoo sandbox uses VirusTotal to perform the experiments, moreover, in the case of VMRay analyzer VTI score is used to label the samples.

To evaluate the proposed research and create the raw and integrated feature set, malware and benign samples were collected from a different source as mentioned in the above section. In order to validate the propose works different portion of samples were taken for validation purpose.

### 3.6 Pre-processing and feature generation

In this stage data were processed and cleaned from noise and irrelevant entries and string information was extracted from the logs file generated by the different sandboxes and features set were constructed. In this research, 34 register hives were identified for malware and 13 register hives for cleanware as shown in the Table 2. Furthermore, these strings were converted into binary feature vector, so they can be given as input to the machine learning algorithm.

## 4. Experimental Results

During the analysis of malware, some locations in the registry and in the Windows file system, have been recognized as important for potential contamination data. Based upon prior work, the following locations were recorded (as illustrated in table 2) (Symantec, 2009; Norton, 2010; Carvey, 2011; Cert-Eu, 2012; RSA, 2013; Bayer et al., 2014; Horsman et al., 2014; Malicious-streams, 2014; Fnal, 2016):

**Table 3: The locations were investigated for digital forensics**

| | **Digital forensics locations** |
|---|---|
| **1** | HKEY_LOCAL_MACHINE\SYSTEM\ControlSet001\Control\Nls\CustomLocale\en-US |
| 2 | HKEY_LOCAL_MACHINE\SYSTEM\ControlSet001\Control\Nls |
| **3** | HKEY_LOCAL_MACHINE\SYSTEM\ControlSet001\Control\SESSION |
| **4** | HKEY_LOCAL_MACHINE\SYSTEM\ControlSet001\Control |
| **5** | HKEY_LOCAL_MACHINE\SYSTEM |
| **6** | HKEY_LOCAL_MACHINE\Software\Microsoft\Rpc |
| **7** | HKEY_LOCAL_MACHINE\SOFTWARE\Wow6432Node\Microsoft\Windows\CurrentVersion |
| **8** | HKEY_LOCAL_MACHINE\SOFTWARE\Microsoft\ |
| **9** | HKEY_LOCAL_MACHINE\SOFTWARE\Wow6432Node\Microsoft\ |
| **10** | HKEY_CURRENT_USER\Software\Microsoft\Windows NT\CurrentVersion\Windows |
| **11** | HKEY_CURRENT_USER\Software\Microsoft\Windows\CurrentVersion\Setup |
| **12** | HKEY_CURRENT_USER\SOFTWARE\Microsoft\Windows\CurrentVersion\Uninstall |
| **13** | HKEY_CURRENT_USER\SOFTWARE\Microsoft\Windows\CurrentVersion\ |
| **14** | HKEY_CURRENT_USER\SOFTWARE\Microsoft\Windows\CurrentVersion\Explorer |
| **15** | Documents and Settings\[user name]\Start Menu\Programs\Startup |
| **16** | %systemdrive%\Documents and Settings\[User Name]\Local Settings\Temp |
| **17** | %Systemdrive%\Users\victim_user\AppData\ |
| **18** | %Systemdrive%\Windows\System32 |
| **19** | %Systemdrive%\Windows\INF\ |
| **20** | %Systemdrive%\Windows\Globalization\Sorting\sortdefault.nls |
| **21** | %Systemdrive%\ |
| **22** | HKEY_LOCAL_MACHINE\software\policies |
| **23** | HKEY_LOCAL_MACHINE\SOFTWARE\Classes\ |
| **24** | HKEY_CURRENT_USER\Software\Microsoft |
| **25** | HKEY_CURRENT_USER\Software\Microsoft\Windows\CurrentVersion\Explorer\Shell Folders |
| **26** | HKEY_LOCAL_MACHINE\Software\Microsoft\Windows\CurrentVersion\explorer\UserShell |
| **27** | HKEY_LOCAL_MACHINE\Software\Microsoft\Windows\CurrentVersion\RunServices |
| **28** | HKEY_CLASSES_ROOT\exefile\shell\open\command |
| **29** | HKEY_CLASSES_ROOT\comfile\shell\open\command |
| **30** | HKEY_LOCAL_MACHINE\Software\CLASSES\batfile\shell\open\command |
| **31** | HKEY_LOCAL_MACHINE\Software\CLASSES\exefile\shell\open\command |
| **32** | HKEY_LOCAL_MACHINE\SOFTWARE\Microsoft\Windows NT\CurrentVersion\Winlogon\Shell |
| **33** | HKEY_LOCAL_MACHINE\Software\Microsoft\Active Setup\Installed Components\KeyName |
| **34** | HKEY_LOCAL_MACHINE\Software\Microsoft\Windows\CurrentVersion\Explorer\ Advanced\Start_ShowDownloads |
| **35** | HKEY_CURRENT_USER\Control Panel\Desktop |
| **36** | HKEY_LOCAL_MACHINE \SOFTWARE\Classes\Interface |
| **37** | HKEY_LOCAL_MACHINE\SOFTWARE\Microsoft\Windows\CurrentVersion\Uninstall\software_name |
| **38** | HKEY_LOCAL_MACHINE\SOFTWARE\Policies\Microsoft\Windows\CurrentVersion\Internet Settings\ZoneMapKey |
| **39** | HKEY_CURRENT_USER\Software\Microsoft\Office\Software_name |
| **40** | HKEY_USERS\% account id %\Software\Adobe\ |
| **41** | HKEY_LOCAL_MACHINE\Software\Classes |
| **42** | HKEY_CLASSES_ROOT\software_name |
| **43** | HKEY_LOCAL_MACHINE\software\microsoft\windows\currentversion\appmanagement\arpcache\ |
| **44** | %Systemdrive%\Users\Public\Documents |
| **45** | %systemdrive%\Program Files\Software_name\ |
| **46** | %SYSTEMDRIVE%\Windows\Fonts |
| **47** | %Systemdrive%\Users\Public\Documents |

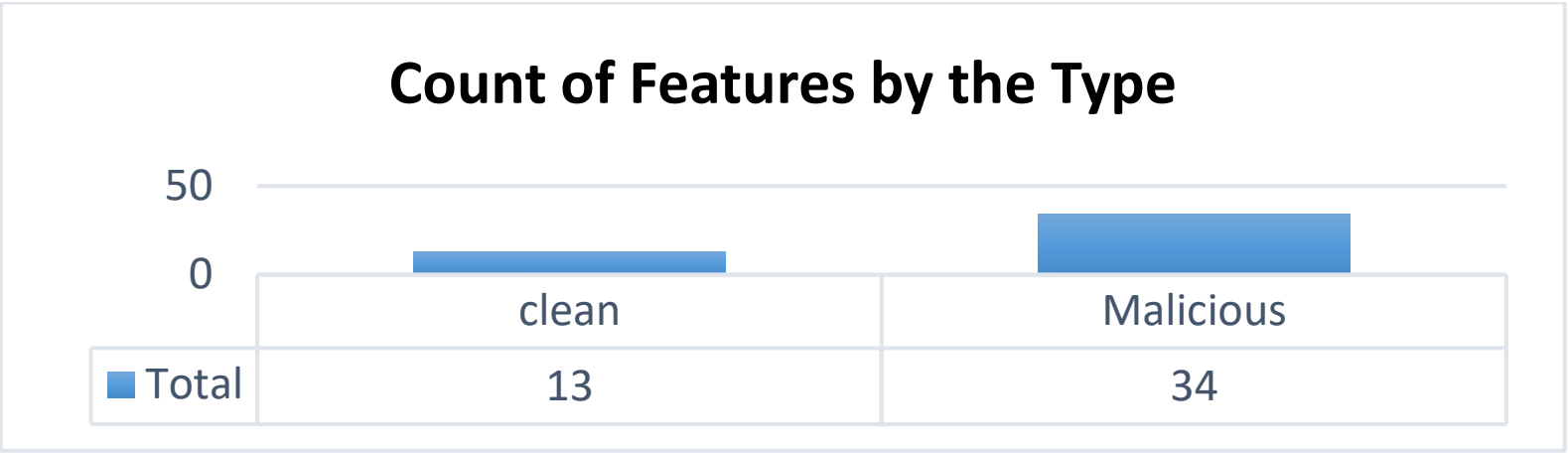

Figure 4: The ratio of clean and malicious hives.

The following sub-sections present an analysis of the findings based on three perspectives:

- The purpose/payload/motivation of the malware (e.g. spying or command and control)
- The type of malware (i.e. Bot, Trojan or Worm)
- The version of the operating system (i.e. Windows 7, 8, or 10)

In each case, the previously identified 47 registry and file locations are analyzed against the 90 samples of both malware and cleanware to proof our initial research question. The raw results from the analysis derive can be found in Appendix A.

## 4.1 Analysis of Malware Motivation

The first analysis is concerned with the research question, whether the motivation of the malware affected the frequency of digital evidence in a particular position within the Registry. As previously identified in Table 2, the types of functionality include:

- Spying and/or steal user data (Trojan)
- Communicating with a control center to receive commands (Botnet)
- Self-Propagation (Worm)
- Benign file

Figure 5- 7 illustrate the degree to which the registry locations are affected. The analysis focussed upon an analysis of the registry against the type of malware. Figure 4(a-c) illustrate the proportion of each type of malware upon the 34 Registry locations. The significant register keys/values for malware and benign values are mentioned below

**Values for Malware**

It has been observed that few hives values are of significant importance when the forensic investigator is looking for malicious activities in the system. The modification of P2, P17, P3, P18 and P1 are higher in proportion as compared to other counterpart, although P17, P18 were also present in Bots and Trojan but other keys impact and modification is higher in malware as compared to them.

- P2 (HKEY_LOCAL_MACHINE\SYSTEM\ControlSet001\Control\Nls ) ,
- P17 - %Systemdrive%\Users\victim_user\AppData\
- P3 -HKEY_LOCAL_MACHINE\SYSTEM\ControlSet001\Control\SESSION
- P18 - %Systemdrive%\Windows\System32
- P1-HKEY_LOCAL_MACHINE\SYSTEM\ControlSet001\Control\Nls\CustomLocale\en-US

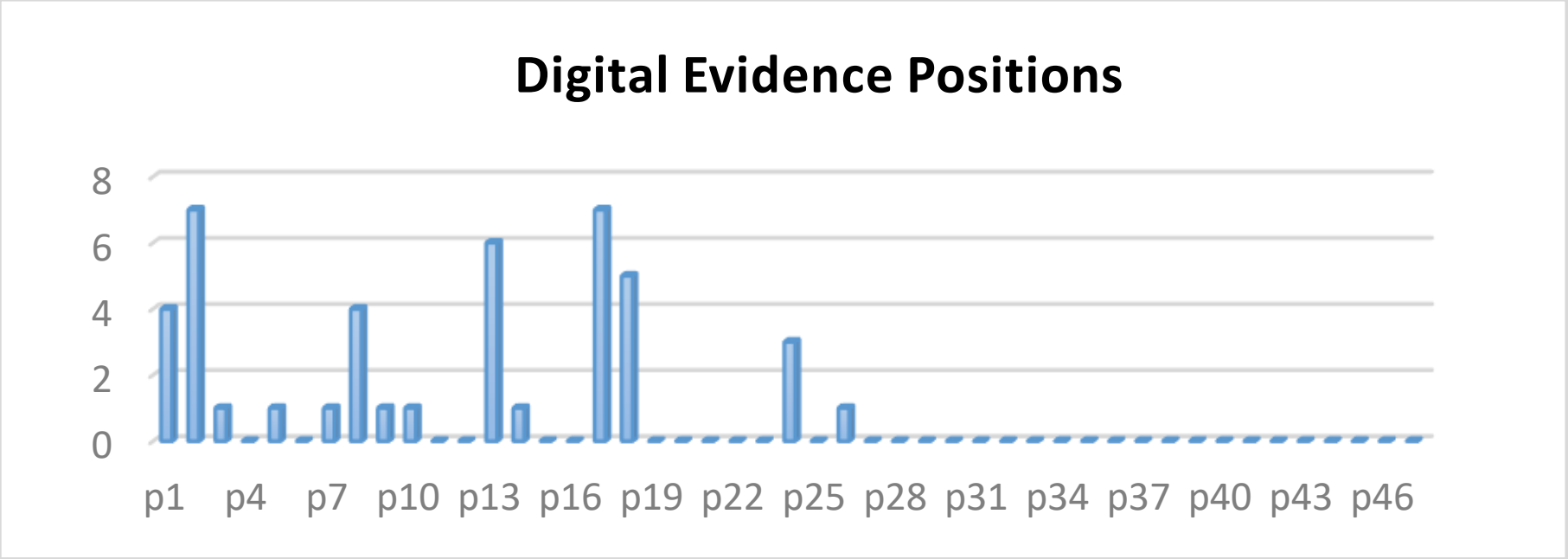

Figure 5: The Impact of Malware on the Registry

An analysis of the Figure 5 and Figure 9 (cleanware) shows that Malware has a slightly different profile in many cases to that of Bots, Trojans, and Worms. For example, Bot has distinctive impacts in the following locations: P1, P2, P8, P13, P17, P18

**Values for Bots**

In the case of bots, the Modification of below-mentioned keys are of indicative of bots activities, the detail of these keys are mentioned below

- P18 – %Systemdrive%\Windows\System32
- P8 - HKEY_LOCAL_MACHINE\SOFTWARE\Microsoft\
- P19 – %Systemdrive%\Windows\INF\
- P17-%Systemdrive%\Users\victim_user\AppData\

- P1-HKEY_LOCAL_MACHINE\SYSTEM\ControlSet001\Control\Nls\CustomLocale\en-US
- P2-HKEY_LOCAL_MACHINE\SYSTEM\ControlSet001\Control\Nls
- P4-HKEY_LOCAL_MACHINE\SYSTEM\ControlSet001\Control

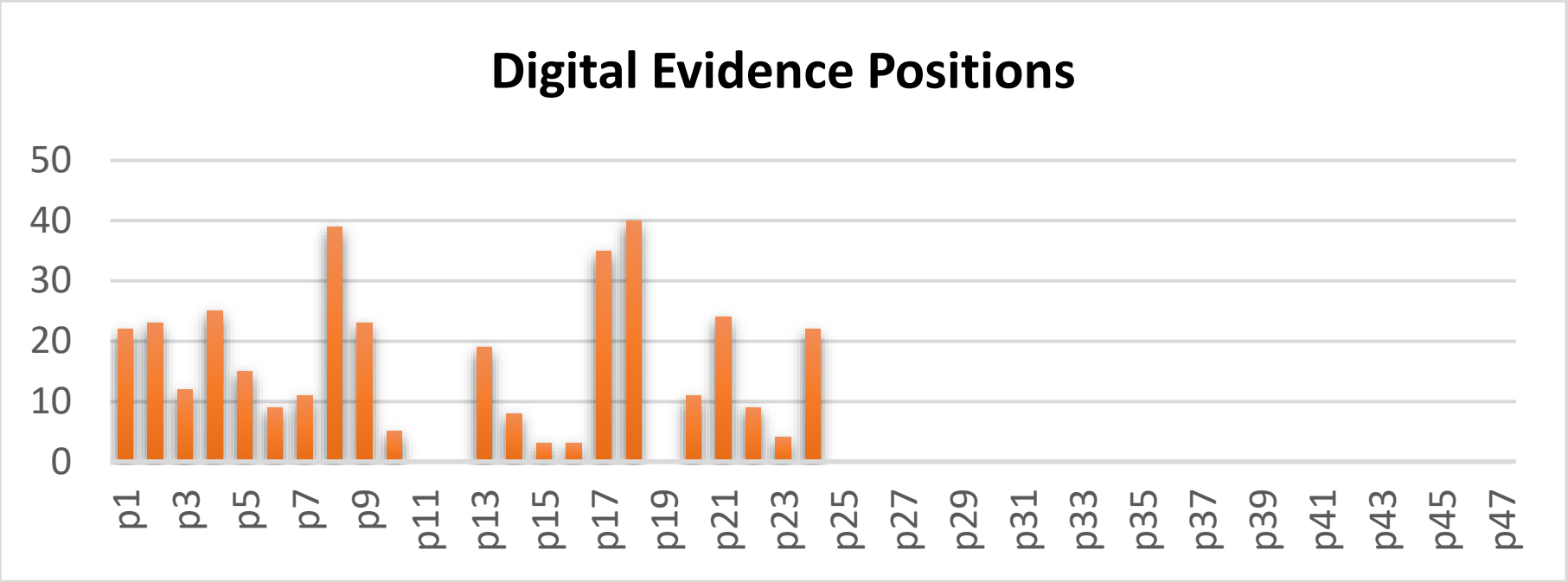

Figure 6: The Impact of Bot on the Registry

**Values for Trojan**

It is notable during analysis that modifications of few keys are higher in the Trojan as compared to malware, the details of these are as follows

- P18 – HKEY_CURRENT_USER\SOFTWARE\Microsoft\Windows\CurrentVersion\Explorer
- P2 - HKEY_LOCAL_MACHINE\SYSTEM\ControlSet001\Control\Nls
- P19 – %Systemdrive%\Windows\INF\
- P17-%Systemdrive%\Users\victim_user\AppData\
- P1-HKEY_LOCAL_MACHINE\SYSTEM\ControlSet001\Control\Nls\CustomLocale\en-US
- P2-HKEY_LOCAL_MACHINE\SYSTEM\ControlSet001\Control\Nls
- P4-HKEY_LOCAL_MACHINE\SYSTEM\ControlSet001\Control
- P20-%Systemdrive%\Windows\Globalization\Sorting\sortdefault.nls

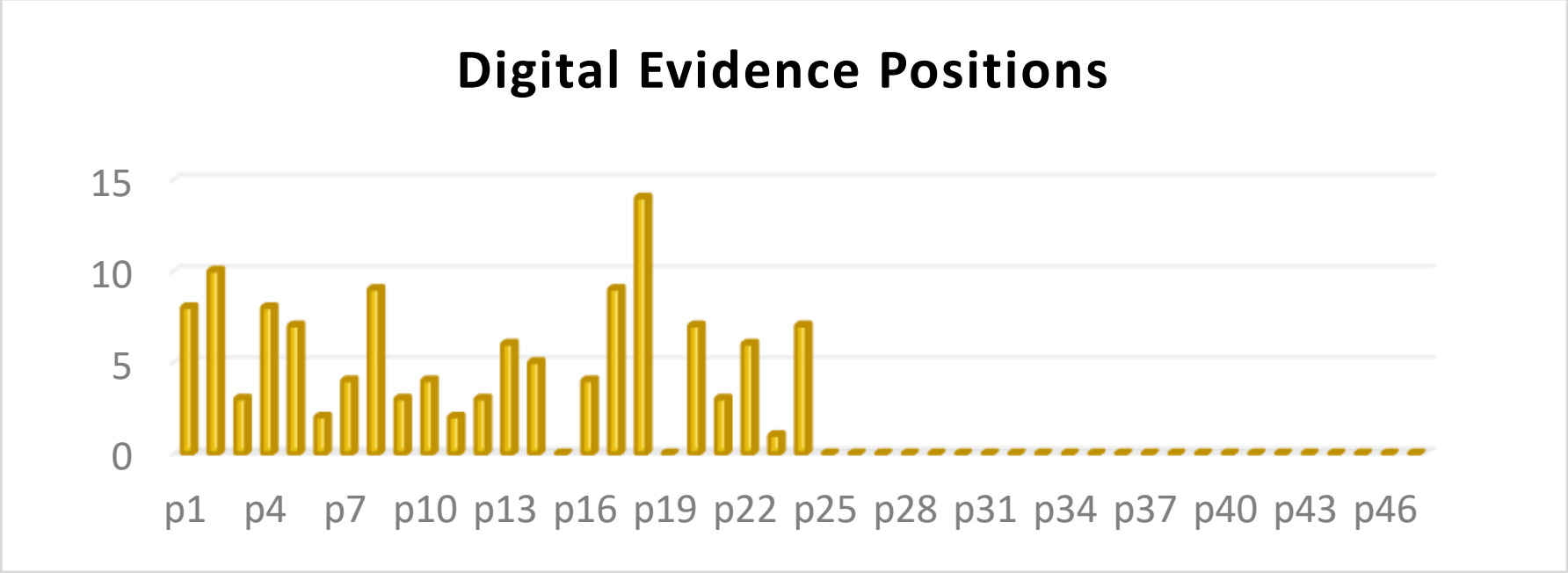

Figure 7: The Impact of Trojan on the Registry

The analysis of the charts shows that Bot and Trojan have similar values. Also the following locations tend to be higher: P18, P19, and P17, P1 in both bots and Trojan.

**Values for Worm**

- P18 – HKEY_CURRENT_USER\SOFTWARE\Microsoft\Windows\CurrentVersion\Explorer
- P17-%Systemdrive%\Users\victim_user\AppData\
- P8-HKEY_LOCAL_MACHINE\SOFTWARE\Microsoft\
- P13-HKEY_CURRENT_USER\SOFTWARE\Microsoft\Windows\CurrentVersion\

- P4-HKEY_LOCAL_MACHINE\SYSTEM\ControlSet001\Control
- P21-%Systemdrive%\
- P20-%Systemdrive%\Windows\Globalization\Sorting\sortdefault.nls

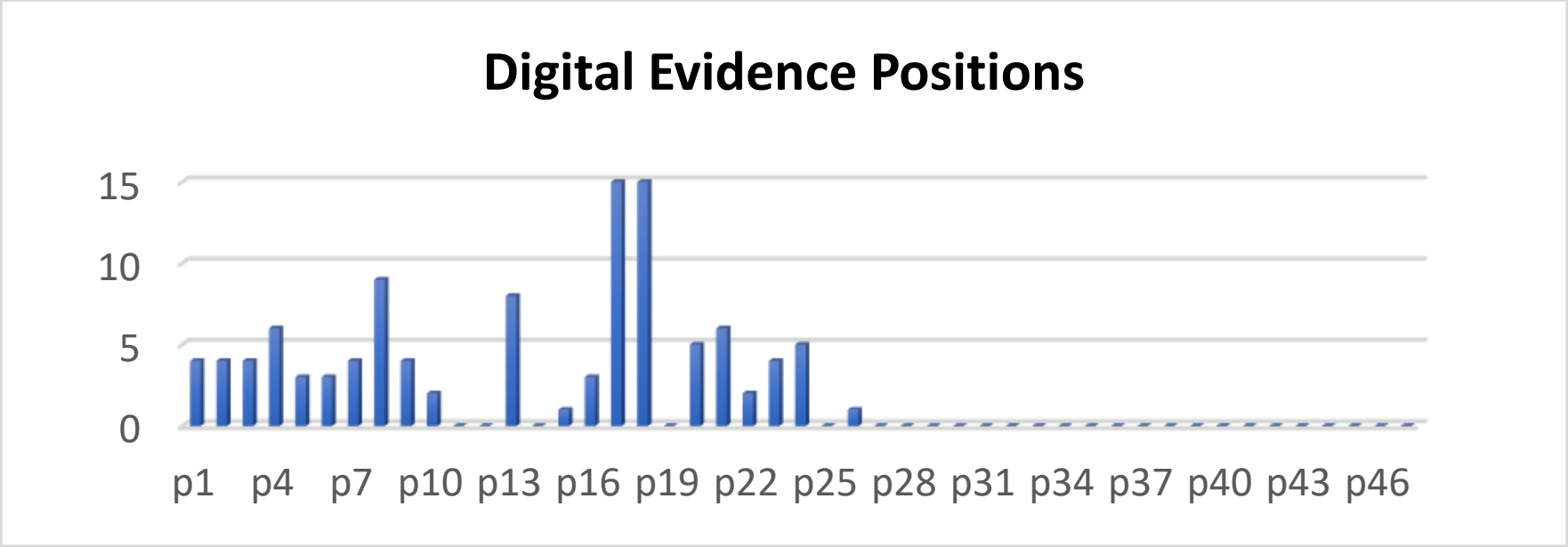

Figure 8: The Impact of worm on the Registry

From the empirical analysis, it has been identified that modification of keys P18 (HKEY_CURRENT_USER\SOFTWARE\Microsoft\Windows\CurrentVersion\Explorer) and P17(-%Systemdrive%\Users\victim_user\AppData\) tends to be higher in all three classes for e.g. Bots, Worms, and Trojans thus we can conclude that these two keys are of great importance for forensic investigator.

**Values for Cleanware**

The distinct thing about this research is that forensic investigator will not only able to find the compromised system on the basis of aforementioned values but he will be also able to distinguish between clean systems if below-mentioned keys will be taken in consideration.

- P45 – %systemdrive%\Program Files\Software_name\
- P38 - HKEY_LOCAL_MACHINE\SOFTWARE\
- Policies\Microsoft\Windows\CurrentVersion\Internet Settings\ZoneMapKey
- P37 – HKEY_LOCAL_MACHINE\SOFTWARE\Microsoft\
- Windows\CurrentVersion\Uninstall\software_name
- P40-HKEY_USERS\% account id %\Software\Adobe\
- P42- HKEY_CLASSES_ROOT\software_name
- P43-HKEY_LOCAL_MACHINE\software\microsoft\windows\
- currentversion\appmanagement\arpcache\
- P44-%Systemdrive%\Users\Public\Documents
- P39-HKEY_CURRENT_USER\Software\Microsoft\Office\Software_name

The modification of these keys will help the forensic investigator to consider system clean instead of malicious without considering the AV scan and test report which will save lots of time as well as resources of the system.

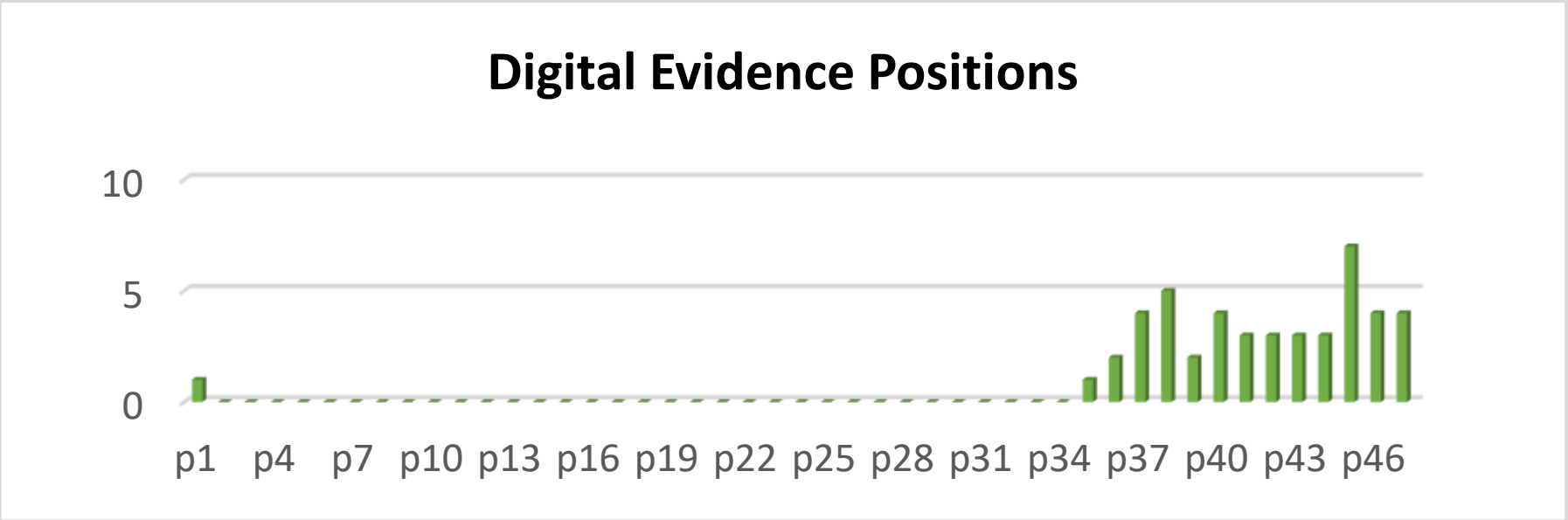

Figure 9: The Impact of cleanware on the Registry

## 4.3 Analysis against Operating System Version

In comparison to the previous two sections where the identification of similarities and differences may assist in helping an investigator in identifying both the presence of malware and the type of payload, the purpose of this comparison is to identify whether any significant differences exist across the last three principal versions of the Windows OS. Notably, as illustrated in Figure 10, 11, 12 the profile exhibited against each OS version is very similar. Going beyond the current state of the art, this study demonstrates that Windows 10 has a very similar impact upon the previously identified 47 registry locations as previous versions.

Upon further examination, there are some small differences that might help identify malware in different versions of the OS.

- Modification to P18 (%Systemdrive%\Windows\System32), P17 (%Systemdrive%\Users\victim_user\AppData\) and P8 (HKEY_LOCAL_MACHINE\SOFTWARE\Microsoft\) has significant impact in all three versions of Windows as shown in figures 10 to 12. It has been observed that register values from P35 to P47 are at lowest level in all three versions of Windows.

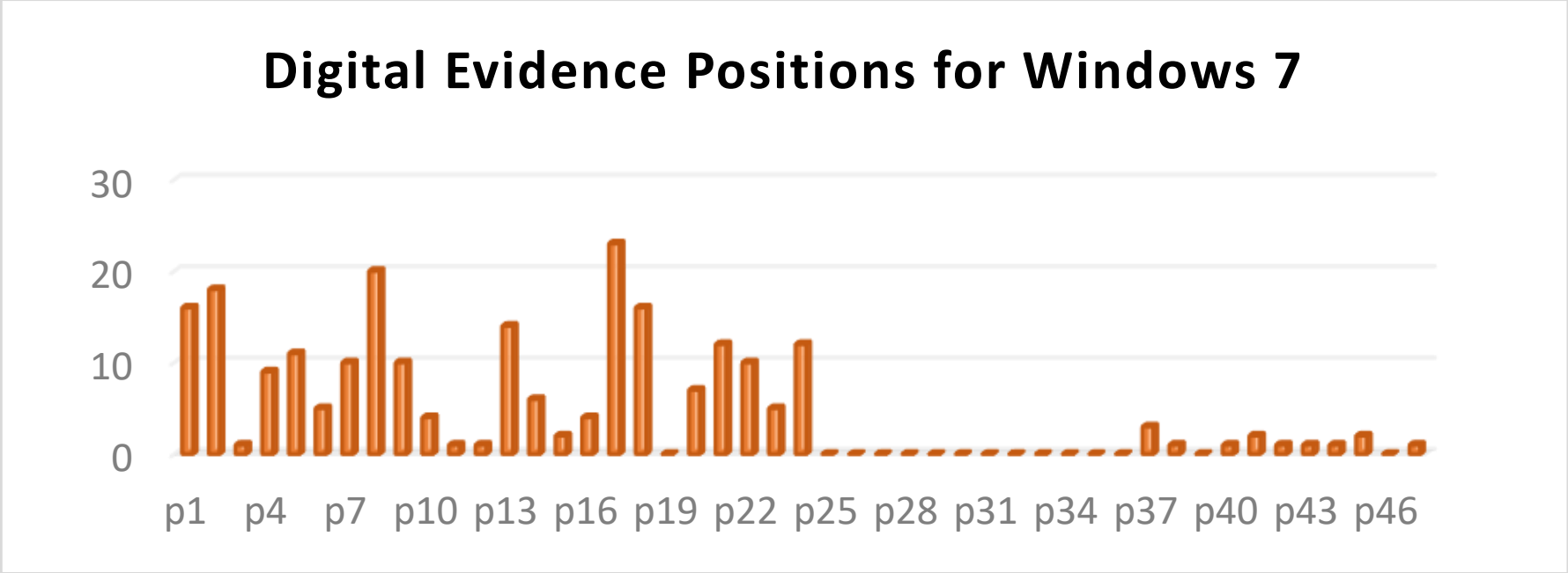

Figure 10: Impact of Malware on the Windows 7 Registry

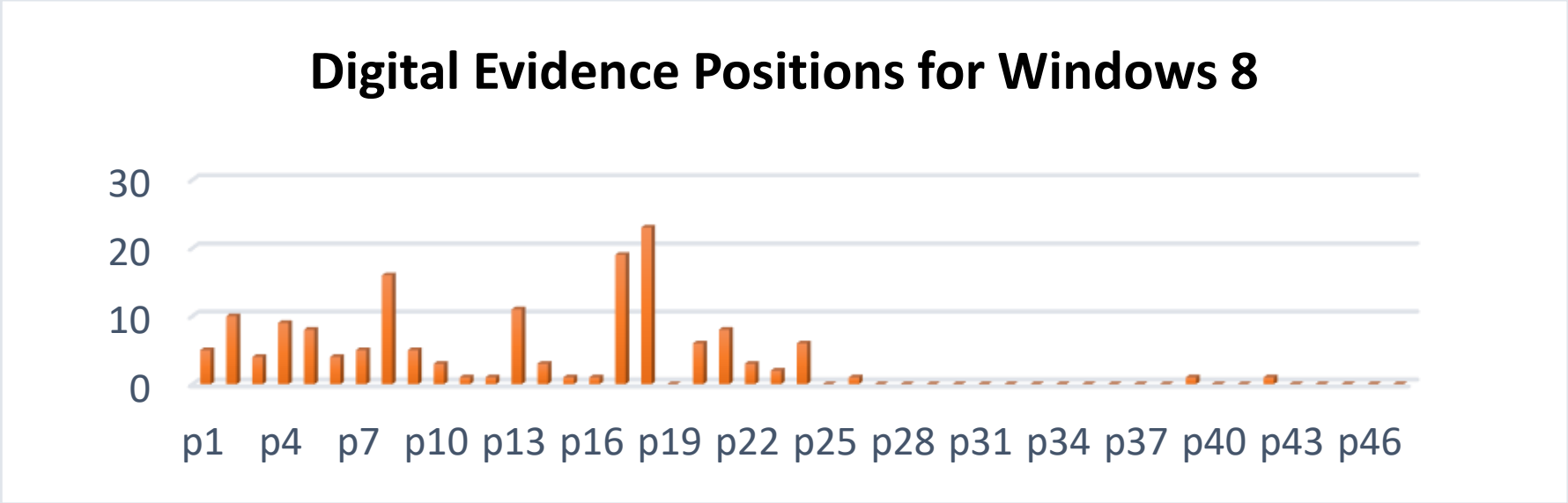

Figure 11: Impact of Malware on the Windows 8 Registry

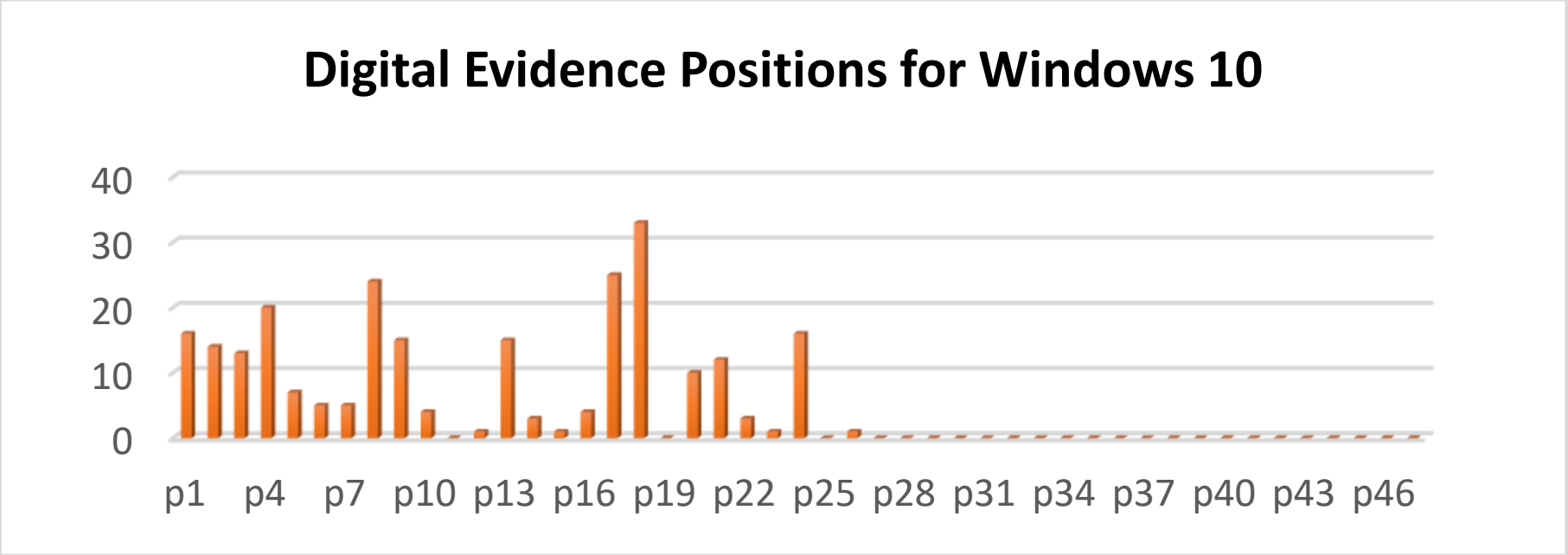

Figure 12: Impact of Malware on the Windows 10 Registry

## 5. Discussion

From the statistical analysis of the results, the main locations of digital assets from malicious and benign software were identified. Furthermore, in each type of software, in each operating system and in each functionality of malware, the most common locations that create digital evidence were recorded. In future research, this analysis could be extended to other categories of malware (Ransomware, Backdoor, etc.) and other forms of functionality.

Whilst there is value to the investigator in better understanding the impact differing forms of malware have upon the different versions of Windows and in particular how the Registry is effected, it would arguably be more useful if this analysis could be applied in a manner that would provide a proactive approach for investigators to be able to detect and classify the type of malware present within a case without having to rely upon AV. This is not designed to replace or remove AV but to complement the approach, particularly in cases where the malware is not being detected by the AV. To this end, an extended experiment was conducted to determine the degree to which the impact upon the registry and file locations is unique to each family of malware (i.e. is it possible given the impact upon the registry and file locations to determine which family the malware belongs).

The results from Appendix A were used, with the 47 locations forming the features from both malware and benign samples. Then a supervised pattern classification approach was selected because they have stronger reliability than unsupervised approaches and a dataset was easily created from existing malware. The samples from the three families were randomized and split into training and testing datasets – with differing proportions to measure the impact that training data has on the overall performance. In this experiment we have used different machine learning algorithm for classification of cleanware and malware classes. The classes that were used are shown in the table below

Table 4: Classes for machine learning

| Class | Label |
|---|---|
| Cleanware | 0 |
| Malware | 1 |
| Worm | -1 |
| Botnet | -2 |
| Trojan | -3 |

During the train, test and validate phases the efficiency and efficacy of the model was measured. Python was utilised and specifically IPython Jupyter notebook v 5.7.2. The Jupyter notebook is an open-source project which is web-based, interactive computing notebook environment which is developed to support data science and scientific computing across the different platform. The first experiment was performed with the label 'Test 1', in which train/test ratio of 80/20 was taken with 47 features, furthermore logic regression, Neural net, Decision tree, Random forest and Boosted tree supervised learning algorithms were utilised and their efficiency and efficacy was measured. It was observed during analysis that Boosted tree algorithm was performing well with 72% accuracy as compared to all other classifiers as depicted in the table 5. Furthermore, investigation was performed by taking training/test ratio of 70/30 to train the model, it had been found that once again Boosted tree outperform all other classifiers with 64% accuracy as shown in table 5. We extend our experiments by taking training/test ratio to 60/40 to see the impact of accuracy on the classifier, we found that decision tree and random forest accuracy increases to 68.4% and 67.5 % respectively, furthermore boosted tree accuracy decrease from 72% to 71%.

From these experiments, it was observed that the best classification accuracy was produced by the Boosted tree with setting 80/20 as compared to other learning algorithms, moreover it was noticed that the Random forest, logistic regression, and decision tree classifiers accuracy increased drastically when we have taken 60/40 ratio but in contrast the accuracy of Boosted tree decrease from 72% to 71% except the neural net whose accuracy remained constant.

**Table 5: Malware Classification Performance**

| Test ID | Train/ Test ratio | Feature tested | Accuracy | | | | |
|---|---|---|---|---|---|---|---|
| | | | Logistic Regression | Neural Net | Decision Tree | Random Forest | Boosted Tree |
| **Test 1** | **80/20** | **47** | 58% | 34% | 62% | 58% | 72% |
| **Test 2** | **70/30** | **47** | 50% | 33.3% | 56% | 64% | 64.9% |
| **Test 3** | **60/40** | **47** | 58% | 33% | 68.4% | 67.5% | 71% |
| **Test 4** | **50/50** | **47** | 64% | 34% | 62% | 62% | 65% |

The results of this extended experiment demonstrate that modeling the impact that malware has upon the registry and hard disk would be a useful approach to detecting the type of malware family. This type of modeling is far faster than traditional AV software and could be applied either as a standalone tool or integrated into existing computer forensic software as an additional forensic analysis. It also has the advantage over AV tools in that, once trained, it does not need to be continually updated to reflect new signatures (which can be hourly for some tools) – merely periodically updated to reflect the general trends in malware composition. Furthermore, the approach could find applications in host-based intrusion detection systems (HIDS) or intrusion protection systems (HIPS) as well as vulnerability scanners.

## 6. Conclusion

The paper has undertaken an investigation into the impact that three core types of malware have upon different versions of the Windows OS – specifically targeting the Registry. Whilst previous research has presented the impact of limited volumes of malware upon the Registry, this is the first study to utilize a large volume of malware across the three core types (Bot, Trojan and Worm) along with clean samples. The results from this analysis largely confirm previous studies but provide a greater granularity as to the impact based on the different types of malware. This study has also extended the prior work by including Windows 10 and evidencing that it has overall a similar impact profile on the Registry as previous versions of the software.

The results have shown that it is possible to accelerate a digital forensics analysis through a preliminary analysis of the registry, the modified timestamps and the use of machine learning or deep learning. Targeting these 47 registry locations can provide a first indication to the digital forensics examiner on whether or not malware is present, but also the type across all common versions of the Windows OS.

This type of analysis has several key advantages over existing approaches: it is faster to scan and identify than AV, it is able to detect and classify new malware prior to AV signatures being developed, it does not need frequent updating and can be built into existing tools with applications in both the forensic and security fields.

**Appendix A.** **Malware Dynamic Analysis Locations (1-34) that forensics artifacts have been recorded during dynamic analysis for each of 180 samples (Malware [p1-p34] and clean [p35-p47]) of worm, Bot and Trojan (table 1,2,& 3),clean (table 4) and for the three Windows operating systems (table 5,6 & 7)**

**Table 1**

| | Worm | | | | | | | | | | | | | | | | | | | | | | | | | | |
|---|---|---|---|---|---|---|---|---|---|---|---|---|---|---|---|---|---|---|---|---|---|---|---|---|---|---|---|
| Locations | P1 | P2 | P3 | P4 | P5 | P6 | P7 | P8 | P9 | P10 | P11 | P12 | P13 | P14 | P15 | P16 | P17 | P18 | P19 | P20 | P21 | P22 | P23 | P24 | P25 | P26 | P27-34 |
| | ● | | | | | | | | | | | | | | | | | | | | | | | | | | |
| | ● | | | | | | | | | | | | | | | | | | | | | | | | | | |
| | ● | | | | | | | | | | | | | | | | | | | | | | | | | | |
| | | | | | | | | | | | | | | | | | | | | | | | | ● | | ● | |
| | | | | | | | | | | | | | ● | | | | ● | | | | | | | | | | |
| | | | | | | | | | | | | | ● | | | | ● | ● | | | | | | | | | |
| | | | | | | | | | | | | | ● | | | | ● | ● | | | | | | | | | |
| | | ● | | | | | | | | | | | | | | | | ● | | | | | | | | | |
| | | | | | | | | | | ● | | | | | | | ● | | | | | | | | | | |
| | | | | | | | | | | | | | | | | | ● | ● | | | | | | | | | |
| | | | | | | | | | | ● | | | | | | | | ● | | | | | | | | | |
| | | | | ● | | | | | | | | | | | | | ● | | | | | | | | | | |
| | ● | ● | | ● | | | ● | ● | | | | | ● | | | | ● | ● | | | | | | | | | |
| | | | | | | | | ● | | | | | | | | | | ● | | ● | | | ● | | | | |
| | | | | | | | | | | | | | | | | | | | | ● | | | ● | | | | |
| | | | | | | | | | | | | | | | | | ● | ● | | ● | | | | | | | |
| | | | | | | | | | | | | | | | | | ● | ● | | ● | | | | | | | |
| | | | | | | | | ● | | | | | | | | | ● | ● | | ● | | | | | | | |
| | | | | ● | ● | ● | ● | ● | ● | | | | ● | | ● | ● | ● | | | | ● | ● | | ● | | | |
| | | | ● | ● | ● | ● | ● | ● | | | | | ● | | | ● | ● | ● | | | ● | | | ● | | | |
| | | | ● | ● | ● | ● | ● | ● | | | | | ● | | | ● | ● | ● | | | ● | | | ● | | | |
| | | ● | | | | | | ● | ● | | | | | | | | ● | ● | | | ● | ● | ● | | | | |

**Table 2**

| Trojan | | | | | | | | | | | | | | | | | | | | | | | | | |
|---|---|---|---|---|---|---|---|---|---|---|---|---|---|---|---|---|---|---|---|---|---|---|---|---|---|
| Locations | P1 | P2 | P3 | P4 | P5 | P6 | P7 | P8 | P9 | P10 | P11 | P12 | P13 | P14 | P15 | P16 | P17 | P18 | P19 | P20 | P21 | P22 | P23 | P24 | P25-34 |
| | | | | | | | | | | ● | | | ● | | | | ● | | | | | | | | |
| | ● | | | | | | | | | ● | | | | | | | ● | ● | | ● | | | | | |
| | | ● | | | | | | | ● | | | | | | | | ● | ● | | ● | | | | | |
| | ● | | | | | | | | | ● | | | | | | | ● | ● | | | ● | | | | |
| | | | | | | | | | | ● | | | | | | | ● | ● | | | | | | | |
| | | ● | | | | | | | | | | | ● | | | | ● | ● | | ● | | | | | |
| | | | | ● | | | | ● | | | | | | | | | ● | ● | | | | | | | |
| | ● | ● | | ● | ● | ● | | | | | | | | | | | | ● | | | ● | ● | | ● | |
| | | | | ● | | | | ● | | | | | | | | | | ● | | | | | | | |
| | | | | | | | | | | | | | | | | ● | | | | | | | | | |
| | ● | ● | | | ● | | ● | ● | ● | | | | ● | ● | | | ● | | | ● | ● | ● | ● | ● | |
| | ● | ● | | ● | ● | ● | ● | ● | | | ● | | ● | ● | | ● | | | | ● | | ● | | ● | |
| | ● | ● | | | | | ● | ● | | | ● | | | ● | | | ● | ● | | | | | | ● | |
| | ● | ● | ● | ● | | | | ● | | | | ● | | ● | | ● | | ● | | ● | | ● | | ● | |
| | ● | ● | ● | ● | | | | ● | | | | ● | | ● | | ● | | ● | | ● | | ● | | ● | |
| | | | | ● | ● | | | ● | | | | ● | ● | | | | | ● | | | | ● | | | |
| | | | ● | ● | ● | | ● | ● | ● | | | | ● | | | | | ● | | | | | | ● | |
| | | | | | ● | | | | | | | | | | | | | | | | | | | | |
| | | | | | ● | | | | | | | | | | | | | ● | | | | | | | |
| | | | ● | | | | | | | | | | | | | | | ● | | | | | | | |
| | | | | | | | | | | | | | | | | | | | | | | | | | |

**Table 1**

| | Bots | | | | | | | | | | | | | | | | | | | | | | | | |
|---|---|---|---|---|---|---|---|---|---|---|---|---|---|---|---|---|---|---|---|---|---|---|---|---|---|
| | P1 | P2 | P3 | P4 | P5 | P6 | P7 | P8 | P9 | P10 | P11 | P12 | P13 | P14 | P15 | P16 | P17 | P18 | P19 | P20 | P21 | P22 | P23 | P24 | P25-34 |
| Locations | | | | | | | ● | | | ● | | | | | | | ● | ● | | | | | | | |
| | | | ● | | | | | ● | | | | | | | | | ● | ● | | | | | | | |
| | | | | ● | ● | | | | | | | | | | | | ● | ● | | | | ● | | | |
| | | | | | | | | | | ● | | | | | | | ● | ● | | | | | | | |
| | | ● | | | | | | | | | | | | | | | ● | ● | | | | | | | |
| | | | | | | | | | | | | | | | | | ● | ● | | | | | | | |
| | | | | | | | | | | ● | | | | | | | ● | ● | | | | | | | |
| | | ● | | | | | | | | | | | | | | | ● | ● | | | | | | | |
| | | | | | | | | | | | | | ● | | | | ● | ● | | | | | | | |
| | | | | ● | | | | | | | | | | | | | ● | ● | | | | | | | |
| | | | | | ● | | | | | | | | | | | | ● | ● | | | | | | | |
| | | | | | | | | | | | | | | | | | ● | ● | | ● | | | | | |
| | ● | ● | | | | | ● | ● | | | | | | | | | | | | ● | | | | ● | |
| | | | | ● | | | | ● | | | | | | | | | | ● | | | | | | | |
| | | | | | ● | | ● | | | | | | ● | | | | ● | ● | | | ● | ● | ● | | |
| | ● | ● | | ● | ● | ● | ● | ● | | | | | ● | | ● | | | | | ● | ● | ● | | ● | |
| | ● | | | | | | | ● | ● | | | | ● | | | | ● | ● | | | ● | | | ● | |
| | | ● | | | | | | ● | ● | | | | | | | | | | | | ● | | | ● | |
| | ● | ● | | | | | ● | ● | | | | | ● | | | | | | | ● | | | | ● | |
| | | | | | | | | ● | ● | | | | | | | | | ● | | | ● | | | | |
| | ● | | ● | ● | | | | ● | ● | | | | ● | | | | | ● | | | ● | | | ● | |
| | | | | | | | | | | | | | | | | | | | | | | | | | |
| | | | | ● | | | | ● | | | | | | | | | | ● | | | | | | | |
| | | | | ● | | | | ● | | | | | | | | | | ● | | | | | | | |
| | | ● | | | | | | ● | | | | | | | | | | | | | | | | | |

**Table 2**

| | Cleanware | | | | | | | | | | | | | |
|---|---|---|---|---|---|---|---|---|---|---|---|---|---|---|
| Locations | P1 | P2-P34 | P36 | P37 | P38 | P39 | P40 | P41 | P42 | P43 | P44 | P45 | P46 | P47 |
| | ● | | | | | | | | | | | ● | ● | ● |
| | | | | | | | ● | | | | | | | |
| | | | | ● | | | | | | | | | | |
| | | | | | ● | | | | | | | | | |
| | | | | | | | | | | | | ● | | |
| | | | | | | | | | ● | | | | | |
| | | | | | | | | | | | | ● | | |
| | | | | | | | | | | | | ● | | |
| | | | | | | | | | | | | ● | | |
| | | | | | | | | | | | | | ● | |
| | | | | | | | | | | | | | | ● |
| | | | | | ● | | | | | | | | | |
| | | | ● | | | | | | | | | | | |
| | | | | ● | | | | | | | | | | |
| | | | | | ● | | | | | | | | | |
| | | | | | | | ● | | | | | | | |
| | | | | | | | | ● | | | | | | |
| | | | | | | | | | | ● | | | | |
| | | | | | | | | | | | ● | | | |
| | | | | | | | | | | | | ● | | |
| | | | | | | | | | | | | | ● | |
| | | | | | | | | | | | | | | ● |
| | | | | | | | | | | | | | | |
| | | | ● | | | | | | | | | | | |
| | | | | ● | | | | | | | | | | |
| | | | | | ● | | | | | | | | | |
| | | | | | | ● | | | | | | | | |
| | | | | | | | ● | | | | | | | |
| | | | | | | | | ● | | | | | | |
| | | | | | | | | | ● | | | | | |
| | | | | | | | | | | ● | | | | |
| | | | | | | | | | | | ● | | | |
| | | | | | | | | | | | | | | ● |
| | | | | | | | | | | | | | ● | |
| | | | | | | | | | | | | ● | | |
| | | | | | | | | | | | ● | | | |
| | | | | | | | | | | ● | | | | |
| | | | | | | | | | ● | | | | | |
| | | | | | | | | ● | | | | | | |
| | | | | | | | ● | | | | | | | |
| | | | | | | ● | | | | | | | | |
| | | | | | ● | | | | | | | | | |
| | | | | ● | | | | | | | | | | |

**Table 3**

## Locations for windows 7

| P1 | P2 | P3 | P4 | P5 | P6 | P7 | P8 | P9 | P10 | P11 | P12 | P13 | P14 | P15 | P16 | P17 | P18 | P19 | P20 | P21 | P22 | P23 | P24 | P25 | P26 | P27 | P28 | P29 | P30 | P31 | P32 | P33 | P34 | P35 | P36 | P37 | P38 | P39 | P40 | P41 | P42 | P43 | P44 | P45 | P46 | P47 |
|---|---|---|---|---|---|---|---|---|---|---|---|---|---|---|---|---|---|---|---|---|---|---|---|---|---|---|---|---|---|---|---|---|---|---|---|---|---|---|---|---|---|---|---|---|---|---|
| • | • | | | | | | | | | | | | | | | | | | | | | | | | | | | | | | | | | | | | | | | | | | | | | |
| • | | | | | | | | | | | | | | | | | | | | | | | | | | | | | | | | | | | | | | | | | | | | | | |
| | | | | | | | | | | | | | | | | | | | | | | | | | | | | | | | | | | | | • | | | | | | | | | | |
| | | | | | | | • | | | | | | | | | | | | | | | | | | | | | | | | | | | | | | | | | | | | | | | |
| | | | | | | | | | | | | | | | | | | | | | | | | | | | | | | | | | | | | | | | | | • | | | | | |
| • | | | | | | | | | | | | | | | | | | | | | | | | | | | | | | | | | | | | | | | | | | | | | | |
| | | | | | | | | | | | | | | | | | | | | | | | | | | | | | | | | | | | | | | | | | | | | • | | |
| | | | | | | | | | | | | • | | | | • | | | | | | | | | | | | | | | | | | | | | | | | | | | | | | |
| | | | | | | | | | | | | • | | | | • | | | | | | | | | | | | | | | | | | | | | | | | | | | | | | |
| | | | | | | | | | | | | • | | | | • | • | | | | | | | | | | | | | | | | | | | | | | | | | | | | | |
| | | | | | | | | | | | | | | | | • | | | | | | | | | | | | | | | | | | | | | | | | | | | | | | |
| | | | | | | | | | | | | | | | | | • | | | | | | | | | | | | | | | | | | | | | | | | | | | | | |
| • | • | | | | | | | | | | | | | | | | | | | | | | | | | | | | | | | | | | | | | | | | | | | | | |
| • | | | | | | | | | | | | | | | | | | | | | | | | | | | | | | | | | | | | | | | | | | | | | | |
| | | | | | | | | | • | | | | | | | • | • | | | | | | | | | | | | | | | | | | | | | | | | | | | | | |
| | • | | | | | | | | | | | | | | | • | • | | | | | | | | | | | | | | | | | | | | | | | | | | | | | |
| | | | | | | | | | | | | | | | | | | | | | | | | | | | | | | | | | | | | • | | | | | | | | | | |
| • | | | | | | | | | • | | | | | | | • | • | | | • | | | | | | | | | | | | | | | | | | | | | | | | | | |
| | | | | | | | | | | | | | | | | | | | | | | | | | | | | | | | | | | | | | | | | • | | | | | | |
| | | | | | | | | | • | | | | | | | | • | | | | | | | | | | | | | | | | | | | | | | | | | | | | | |
| | | | | | | | | | | | | | | | | | | | | | | | | | | | | | | | | | | | | | | | | | | | | • | | |
| | • | | | | | | | | | | | | | | | • | • | | | | | | | | | | | | | | | | | | | | | | | | | | | | | |
| | | | | | | | | | | | | • | | | | | | | | | | | | | | | | | | | | | | | | | | | | | | | | | | |
| | | | | | | | | | | | | • | | | | | | | | | | | | | | | | | | | | | | | | | | | | | | | | | | |
| | | | | | | | | | | | | | | | | | | | | | | | | | | | | | | | | | | | | • | | | | | | | | | | |
| | | | • | | | | | | | | | | | | | • | • | | | | | | | | | | | | | | | | | | | | | | | | | | | | | |
| | | | | | | | | | | | | | | | | | | | | | | | | | | | | | | | | | | | | | | | • | | | | | | | |
| | | | | | | | | | | | | | | | | | • | | | | | | | | | | | | | | | | | | | | | | | | | | | | | |

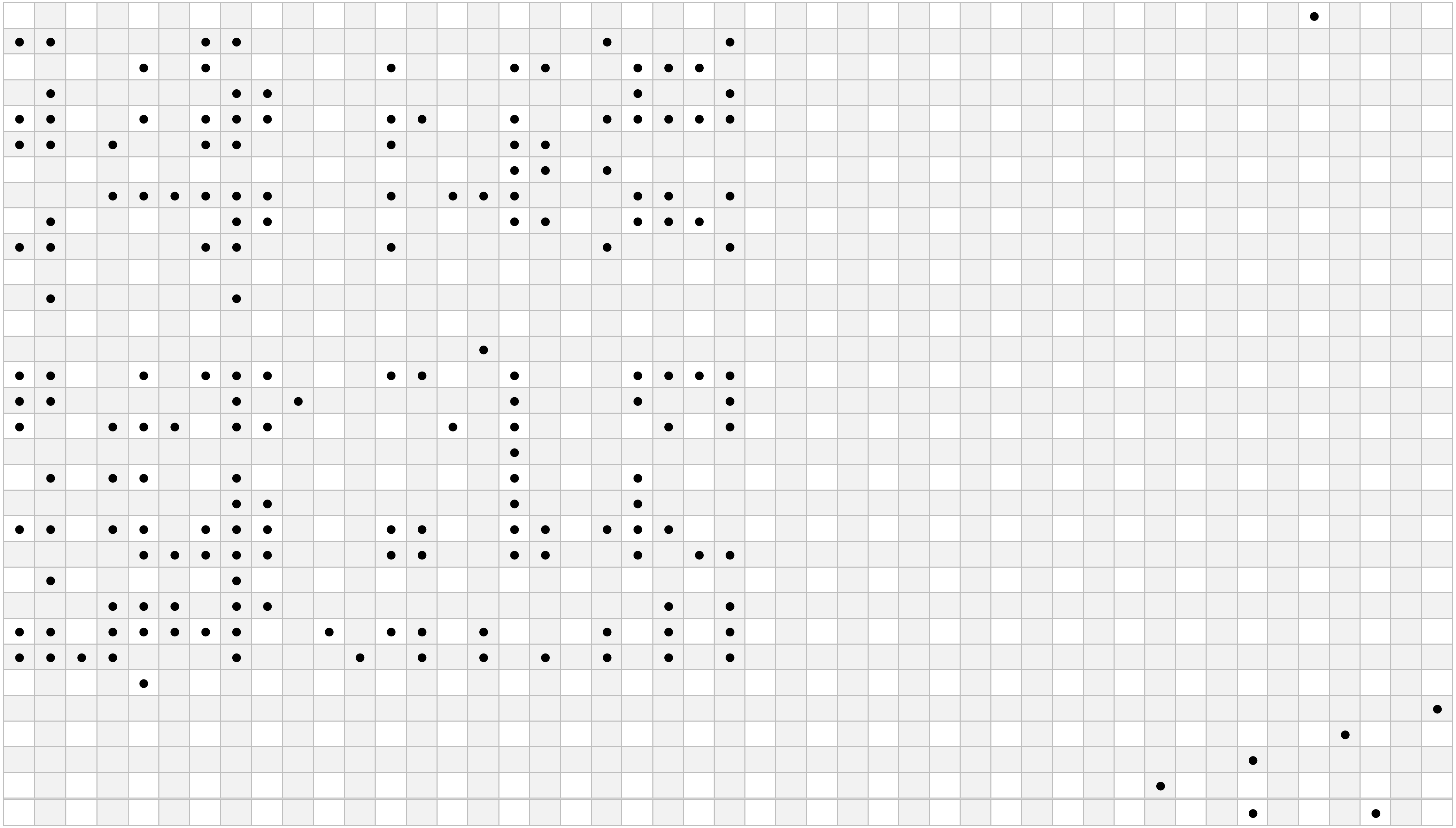

**Table 6**

| Locations for windows 8 | | | | | | | | | | | | | | | | | | | | | | | | | | | | | | | | | | | | | | | | | | | | | | |
|---|---|---|---|---|---|---|---|---|---|---|---|---|---|---|---|---|---|---|---|---|---|---|---|---|---|---|---|---|---|---|---|---|---|---|---|---|---|---|---|---|---|---|---|---|---|---|
| P1 | P2 | P3 | P4 | P5 | P6 | P7 | P8 | P9 | P10 | P11 | P12 | P13 | P14 | P15 | P16 | P17 | P18 | P19 | P20 | P21 | P22 | P23 | P24 | P25 | P26 | P27 | P28 | P29 | P30 | P31 | P32 | P33 | P34 | P35 | P36 | P37 | P38 | P39 | P40 | P41 | P42 | P43 | P44 | P45 | P46 | P47 |
| | | | | | | | | | | | | | | | | | | | | | | | ● | | | | | | | | | | | | | | | | | | | | | | | |
| | | | | | | ● | | | ● | | | | | | | ● | ● | | | | | | | | | | | | | | | | | | | | | | | | | | | | | |
| | | | | | | | | | | | | | | | | | | | | | | | ● | | ● | | | | | | | | | | | | | | | | | | | | | |
| | | | | | | | | | | | | ● | ● | | | | | | | | | | | | | | | | | | | | | | | | | | | | | | | | | |
| | | | | | | | ● | | | | | ● | | | | | | | | | | | | | | | | | | | | | | | | | | | | | | | | | | |
| | | | | | | | | | | | | ● | | | | ● | ● | | | | | | | | | | | | | | | | | | | | | | | | | | | | | |
| | | | | | | | | | ● | | | ● | | | | ● | | | | | | | | | | | | | | | | | | | | | | | | | | | | | | |
| | ● | | | | | | | | | | | | | | | | | | | | | | | | | | | | | | | | | | | | | | | | | | | | | |
| | | | | | | | | | | | | | | | | ● | ● | | | | | | | | | | | | | | | | | | | | | | | | | | | | | |
| | | | | | | | | | ● | | | | | | | ● | | | | | | | | | | | | | | | | | | | | | | | | | | | | | | |
| | | | ● | | | | ● | | | | | | | | | | ● | | | | | | | | | | | | | | | | | | | | | | | | | | | | | |
| ● | ● | | ● | ● | ● | ● | ● | | | | | ● | | ● | | | | | ● | ● | ● | | ● | | | | | | | | | | | | | | | | | | | | | | | |
| | | | ● | | | | ● | | | | | | | | | | ● | | | | | | | | | | | | | | | | | | | | | | | | | | | | | |
| ● | ● | | | ● | | ● | ● | | | | | ● | | | | ● | | | | | | | | | | | | | | | | | | | | | | | | | | | | | | |
| | | | | | | | ● | | | | | | | | | | ● | | ● | | | ● | | | | | | | | | | | | | | | | | | | | | | | | |
| | | | | | | | | | | | | | | | | ● | ● | | ● | | | | | | | | | | | | | | | | | | | | | | | | | | | |
| | | ● | ● | ● | ● | ● | ● | | | | | ● | | | ● | ● | ● | | | ● | | | ● | | | | | | | | | | | | | | | | | | | | | | | |
| | ● | ● | | | | | ● | ● | | | | | | | | ● | ● | | | ● | | ● | | | | | | | | | | | | | | | | | | | | | | | | |
| | | | | | | | ● | ● | | | | | | | | | ● | | | ● | | | | | | | | | | | | | | | | | | | | | | | | | | |
| | | | ● | | | | ● | | | | | | | | | | ● | | | | | | | | | | | | | | | | | | | | | | | | | | | | | |
| | ● | ● | | | | | | | | | | | | | | ● | ● | | | ● | ● | | | | | | | | | | | | | | | | | | | | | | | | | |
| ● | ● | | ● | ● | ● | | ● | ● | | | | | | | | ● | ● | | ● | ● | | | | | | | | | | | | | | | | | | | | | | | | | | |
| ● | ● | | | | | | ● | ● | | | | ● | ● | | | ● | ● | | ● | ● | | | | | | | | | | | | | | | | | | | | | | | | | | |
| | | | ● | | | | ● | ● | | | | | | | | ● | ● | | | ● | | | ● | | | | | | | | | | | | | | | | | | | | | | | |
| | ● | ● | | | | | | | | | | | | | | | ● | | | | | | | | | | | | | | | | | | | | | | | | | | | | | |
| | | | ● | ● | ● | | ● | | | | | | | | | | | | | | | | | | | | | | | | | | | | | | | | | | | | | | | |
| ● | ● | | | | | ● | ● | | | ● | | | ● | | | ● | ● | | | | | | ● | | | | | | | | | | | | | | | | | | | | | | | |
| | | | ● | ● | | | ● | | | | ● | ● | | | | | ● | | | | ● | | | | | | | | | | | | | | | | | | | | | | | | | |
| | | | | ● | | | | | | | | | | | | | ● | | | | | | | | | | | | | | | | | | | | | | | | | | | | | |

**Table 7**

**Locations for windows 10**

| P1 | P2 | P3 | P4 | P5 | P6 | P7 | P8 | P9 | P10 | P11 | P12 | P13 | P14 | P15 | P16 | P17 | P18 | P19 | P20 | P21 | P22 | P23 | P24 | P25 | P26 | P27 | P28 | P29 | P30 | P31 | P32 | P33 | P34 | P35 | P36 | P37 | P38 | P39 | P40 | P41 | P42 | P43 | P44 | P45 | P46 | P47 |
|---|---|---|---|---|---|---|---|---|---|---|---|---|---|---|---|---|---|---|---|---|---|---|---|---|---|---|---|---|---|---|---|---|---|---|---|---|---|---|---|---|---|---|---|---|---|---|
| | ● | | | | | ● | ● | ● | | | | | | | | | | | | | | | | | | | | | | | | | | | | | | | | | | | | | | |
| | | | | | | | | | | | | | | | | | | | | | | | ● | | | | | | | | | | | | | | | | | | | | | | | |
| | | | | | | | | | | | | | | | | ● | ● | | | | | | ● | | ● | | | | | | | | | | | | | | | | | | | | | |
| ● | | | | | | | | | | | | | | | | | | | | | | | | | | | | | | | | | | | | | | | | | | | | | | |
| | | | | | | | ● | | | | | | | | | | | | | | | | | | | | | | | | | | | | | | | | | | | | | | | |
| | | ● | | | | | | | | | | | | | | | | | | | | | | | | | | | | | | | | | | | | | | | | | | | | |
| | | ● | | | | | ● | | | | | | | | | ● | ● | | | | | | | | | | | | | | | | | | | | | | | | | | | | | |
| | | | ● | ● | | | | | | | | | | | | ● | ● | | | | ● | | | | | | | | | | | | | | | | | | | | | | | | | |
| | ● | | | | | | | | | | | | | | | | ● | | | | | | | | | | | | | | | | | | | | | | | | | | | | | |
| ● | | | | | | | | | | | | ● | | | | ● | | | | | | | | | | | | | | | | | | | | | | | | | | | | | | |
| | ● | | | | | | | | | | | | | | | ● | | | | | | | | | | | | | | | | | | | | | | | | | | | | | | |
| | | | | | | | | | ● | | | | | | | ● | | | | | | | | | | | | | | | | | | | | | | | | | | | | | | |
| ● | | | | | | | | | ● | | | | | | | ● | ● | | ● | | | | | | | | | | | | | | | | | | | | | | | | | | | |
| | ● | | | | | | | ● | | | | | | | | ● | ● | | ● | | | | | | | | | | | | | | | | | | | | | | | | | | | |
| | | | | | | | | | | | | | | | | ● | ● | | | | | | | | | | | | | | | | | | | | | | | | | | | | | |
| | | | | | | | | | ● | | | | | | | ● | ● | | | | | | | | | | | | | | | | | | | | | | | | | | | | | |
| | ● | | | | | | | | | | | | | | | | | | | | | | | | | | | | | | | | | | | | | | | | | | | | | |
| | 2 | | | | | | | | ● | | | | | | | ● | ● | | | | | | | | | | | | | | | | | | | | | | | | | | | | | |
| | | | | | | | | | | | | | | | | ● | ● | | | | | | | | | | | | | | | | | | | | | | | | | | | | | |
| | ● | | | | | | | | | | | ● | | | | ● | ● | | ● | | | | | | | | | | | | | | | | | | | | | | | | | | | |
| | | | ● | | | | ● | | | | | | | | | ● | ● | | | | | | | | | | | | | | | | | | | | | | | | | | | | | |